\documentclass[useAMS,usenatbib]{mn2e}
\usepackage{epsfig}
\usepackage{rotating}

\def\lesssim{\,\lower2truept\hbox{${<\atop\hbox{\raise4truept\hbox{$\sim$}}}$}\,}
\def\gtrsim{\,\lower2truept\hbox{${>\atop\hbox{\raise4truept\hbox{$\sim$}}}$}\,}

\title[Foreground removal for SKA EoR with the CCA]{Foreground removal for Square Kilometre Array observations of the Epoch of Reionization with the Correlated Component Analysis}
\author[Bonaldi et al.]{
\parbox[t]{\textwidth}
{Anna Bonaldi$^{1}$\thanks{E-mail: anna.bonaldi@manchester.ac.uk}, Michael L. Brown$^{1}$}
\vspace*{8pt} \\
$^{1}$Jodrell Bank Centre for Astrophysics, School of Physics \& Astronomy, University of Manchester, Oxford Road, Manchester M13 9PL, U.K.}

\begin{document}

\date{}

\pagerange{\pageref{firstpage}--\pageref{lastpage}} \pubyear{2012}

\maketitle

\label{firstpage}

\begin{abstract}
We apply the Correlated Component Analysis (CCA) method on simulated data of the Square Kilometre Array, with the aim of accurately cleaning the 21\,cm reionization signal from diffuse foreground contamination. The CCA has been developed for the Cosmic Microwave Background, but the application of the Fourier-domain implementation of this method to the reionization signal is straightforward. 

The CCA is a parametric method to estimate the frequency behaviour of the foregrounds from the data by using second-order statistics. We test its performance on foreground simulations of increasing complexity, designed to challenge the parametric models adopted. We also drop the assumption of spectral smoothness that most of the methods rely upon. We are able to clean effectively the simulated data  across the explored frequency range (100--200\,MHz) for all the foreground simulations. This shows that the CCA method is very promising for EoR component separation.  

\end{abstract}
\begin{keywords}
cosmology: theory -- dark ages, reionization, first stars -- diffuse radiation -- radio continuum: ISM -- methods: data analysis  
\end{keywords}

\section{Introduction}\label{sect:intro}
The epoch of reionization (EoR) is one of the last unobserved eras of our Universe. During this era, the radiation emitted by the first cosmic structures causes the Universe to go from neutral to fully ionised. The study of the EoR thus provides invaluable information on the early stages of structure formation. The current constraints on reionization from quasar spectra \citep{fan2006,becker2013} and the Cosmic Microwave Background \citep[CMB,][]{wmap9-cosmo,zahn2012,planck-cosmo} imply a complex reionization history, going on from $z=20$--30 to $z=6$. The most promising way to probe the EoR is through the observation of the 21\,cm hyperfine transition of neutral hydrogen \citep[see e.g.,][for a review]{furlanetto2006, morales2010, pritchard2012}. 

A number of projects are currently under way to detect the 21\,cm signal from reionization: the Low Frequency Array (LOFAR)\footnote{http://www.lofar.org}; the Giant Metrewave Radio Telescope (GMRT)\footnote{http://gmrt.ncra.tifr.res.in}; the Murchison Widefield Array (MWA)\footnote{http://www.mwatelescope.org}; the Precision Array to Probe the Epoch of Reionization (PAPER)\footnote{http://astro.berkeley.edu/dbacker/eor}, the 21 Centimeter Array (21CMA)\footnote{21cma.bao.ac.cn}. Such instruments aim at a statistical detection of the signal and are currently providing the first upper limits \citep{paciga2013,dillon2013,parsons2013}. The next generation instruments, like the Square Kilometer Array \citep[SKA,\footnote{http://www.skatelescope.org}][]{mellema2013} will be able to accurately map this radiation over a wide redshifts range.  

One of the most difficult aspects of the 21\,cm measurement is the presence foreground emission, due to our Galaxy and extragalactic sources, which is about four orders of magnitude brighter than the cosmological signal \citep{jelic2008,jelic2010,bernardi2010,yatawatta2013,moore2013}. 
Most approaches envisage a first step, where  bright point-like sources are subtracted \citep{dimatteo2002,dimatteo2004} and a second one, which deals with the diffuse components. The latter can be done by fitting the foregrounds in frequency by assuming that they are spectrally smooth \citep{oh2003,santos2005,gleser2008,jelic2008,harker2009,liu2009,bernardi2010,liu2011,petrovic2011,liu2012,dillon2013,moore2013}. 

Alternatively, blind methods have been proposed, which do not rely on hypotheses regarding the foreground spectra \citep{chapman2012,chapman2013}. In fact, a concern related to the spectral fitting approach is the ability to correctly model and accurately fit the spectra \citep[see e.g.][for the effect of incorrect modelling and fitting]{morales2006}. 

In this work we adapt to the EoR data the Correlated Component Analysis  \citep[CCA,][]{bonaldi2006,ricciardi2010} method. The CCA is a ``model learning'' algorithm, which estimates the frequency spectrum of the foreground components from the data exploiting second-order statistics. This method can be referred to as semi-blind, as it exploits some previous knowledge of the foreground emission but it estimates the relevant information from the data. As such, it falls somewhere in between the two categories of approaches outlined above. 

The main motivation for introducing this approach is for its ability to improve the understanding of the foregrond components. For CMB studies, for which this method has been originally developed, the CCA has been successfully used to improve the modelling of the poorly known anomalous microwave emission \citep{bonaldi2007, special, gouldbelt}. For the current application, some outstanding questions are the smoothness of the foreground components at the relevant frequencies and scales and which are the best spectral models to describe them. The ability to test our hypotheses regarding the spectral properties of the foregrounds and to refine our spectral models will be a crucial prerequisite for the application of all parametric foreground removal approaches. For this reason, we test the CCA method on foreground simulations of increasing complexity. For a direct comparison of the CCA method with other EoR foreground-removal methods we refer the reader to \cite{chapman_SC}.
%
%

The paper is organised as follows: in Section \ref{sec:one} we describe the CCA method; in Section \ref{sec:two} we describe the simulations that we use to test our analysis. In Section \ref{sec:analysis} we describe in details the analysis performed; In Section~\ref{sec:results} we assess the quality of the foreground cleaning on maps and power spectra. Finally we present our conclusions in Section \ref{sec:conclu}.

\section{The Correlated Component Analysis}\label{sec:one}

\subsection{Fourier-domain CCA}\label{sec:method}
For component separation purposes, it is convenient to model the data as a linear mixture of the components. In this work we apply the linear mixture model to the Fourier domain. This is the most natural choice for interferometric data, since measurements are performed directly in the $uv$ plane. What follows is a brief description of the Fourier-domain implementation of the CCA method \citep[for more details see][]{ricciardi2010}.

For each point in the transformed $uv$ plane we write the data model as  
\begin{equation}
\bmath{x}=\bmath{\sf  B}\bmath{\sf  H}\bmath{s}+\bmath{n}\label{modhcca}.
\end{equation}
The vectors $\bmath{x}$ and $\bmath{n}$ have dimension $N_{\rm f}$ (number of frequency channels) and contain the data and the noise in the Fourier space, respectively. The vector $\bmath{s}$ has dimension $N_{\rm c}$ (number of components) and contains the astrophysical emission components; the diagonal $N_{\rm f}\times N_{\rm f}$ matrix $\bmath{\sf B}$ contains the instrumental beams in Fourier space and the  $N_{\rm f}\times N_{\rm c}$ matrix $\bmath{\sf H}$, called the mixing matrix, contains the intensity of the components at all frequencies. The mixing matrix is the key ingredient of component separation: if $\bmath{\sf H}$ is known, the problem reduces to a suitable inversion of eq.~(\ref{modhcca}). Unfortunately, the mixing matrix is in general not known, at least not to the precision required for an accurate component separation. This led to the development of methods to estimate the mixing matrix from the data. 

The CCA is one of these methods. It exploits the correlation of the data between different frequencies to estimate the mixing matrix. The additional assumptions made by the CCA are that the mixing matrix is constant within the considered area of the sky, and that its unknown elements can be reduced by adopting a suitable parametrization $\bmath{\sf  H}=\bmath{\sf  H}(\bmath{p})$. We will come back to these assumptions in Sect.~\ref{sec:analysis}.

Starting from the linear mixture data model, and assuming that $\bmath{\sf  H}$ is constant for all the points in the $uv$ plane, we can easily derive the following relation between the cross-spectra of the data $\bmath{\sf  C}_{\bmath{x}}(k)$, sources $\bmath{\sf  C}_{\bmath{s}}(k)$ and noise, $\bmath{\sf  C}_{\bmath{n}}(k)$, all depending on the Fourier mode $k$:
\begin{equation}
\bmath{\sf C}_{\bmath{x}}(k)=\bmath{\sf B}(k)\bmath{\sf H}\bmath{\sf C}_{\bmath{s}}(k)\bmath{\sf H}^{\rm T}\bmath{\sf B}^\dagger(k)+\bmath{\sf C}_{\bmath{n}}(k),
\label{hcca_constr}
\end{equation}
where the dagger superscript denotes
the adjoint matrix. 

If $\bmath{x}(i,j)$ is the two-dimensional discrete Fourier transform of the data on a planar grid, the power spectrum can be obtained as the average of $\bmath{x}(i,j) \bmath{x}^{\dag}(i,j)$ in annular bins $D_{k}$, $k=1,\ldots,k_{\rm max}$:
\begin{equation}
\bmath{\sf C}_{\bmath{x}}(k) = \frac{1}{M_{k}}
\sum_{i,j 
\in D_{k}} \bmath{x}(i,j)  \bmath{x}^{\dag}(i,j), \label{dataspectrum}
\end{equation}
where $M_{k}$ is the number of pairs $(i,j)$ contained in the spectral  bin denoted by $D_{k}$. The minimum and maximum $k$ for the spectral bins depend on the area of the sky considered and on the instrumental resolution, respectively. Since the foreground spectra are a smooth function of $k$, the number of bins has little effect on the results.

To write the likelihood in a compact form, we define vectors containing all the elements of the matrices $\bmath{\sf C}$ in eq.~(\ref{hcca_constr}) for all frequencies/components and for a set of spectral bins $k$. If $\bmath{d}$ contains the elements of ${\bmath{\sf C}}_{\bmath{x}}(k) - {\bmath{\sf C}}_{\bmath{n}}(k)$ and $\bmath{c}$ contains the elements of ${\bmath{\sf C}}_{\bmath{s}}(k)$, eq.~(\ref{hcca_constr}) becomes 
\begin{equation}
\bmath{d} = \bmath{\sf H}_{kB}\bmath{c}+ \bmath{\epsilon},\label{fd_cca_error}
\end{equation}
where $\bmath{\sf H}_{kB}$ contains the elements of the Kronecker product $[\bmath{\sf B}(k)\bmath{\sf H}]\otimes[\bmath{\sf B}(k)\bmath{\sf H}]$ and $\bmath{\epsilon}(k)$ represents the error on the noise power spectrum. 

The unknowns (the parameter vector $\bmath{p}$ describing the mixing matrix and the source cross-spectra $\bmath{c}$) are finally obtained by minimizing the functional
\begin{eqnarray}\label{hcca_objective}
\!\!&&\bmath{\Phi}[\bmath{p},\bmath{c}]=\\
\!\!&&\!\!\!\![\bmath{d}-\bmath{\sf H}_{kB}(\bmath{p})\cdot \bmath{c}]^T \bmath{\sf N}_{\epsilon}^{-1} [\bmath{d}-\bmath{\sf H}_{kB}(\bmath{p})\cdot \bmath{c}]+\lambda\bmath{c}^T\bmath{C}\bmath{c} \nonumber 
\end{eqnarray}
where the diagonal matrix $\bmath{\sf N}_{\epsilon}$ contains the covariance of the noise error $\bmath{\epsilon}$. The CCA method also provides an estimate of the statistical errors on the parameters (see Ricciardi et al. 2010 for more details).

The quadratic form in eq.~(\ref{hcca_objective}) represents the log-likelihood for $\bmath{p}$ and $\bmath{c}$. The term $\lambda\mathbf{c}^T\mathbf{C}\mathbf{c}$ is a quadratic stabilizer for the source power cross-spectra, where the matrix $\mathbf{C}$ must be suitably chosen.  This term can be viewed as a log-prior density for the source power cross-spectra, with the parameter $\lambda$ being tuned to balance the effects of data fit and regularization. However, in a high signal-to-noise case such as the one considered here, there is no need for such regularization, so in this work we have used $\lambda=0$.

The minimization in eq.~(\ref{hcca_objective}) is perfomed with the ``simulated annealing'' (SA) method. This algorithm employs a random search which not only accepts changes that decrease the objective function $\bmath{\Phi}$, but also some changes that increase it. The latter are accepted with a probability which depends on $\Delta \bmath{\Phi}$ and on a control parameter, known as the system ``temperature'', decreasing as the minimization proceeds. The major advantage of SA over other methods is its ability to span the entire parameter range and avoid becoming trapped at local minima.

\subsection{Application to 21\,cm data}\label{sec:himodel}
Essentially, the linear mixture data model with constant mixing matrix postulates that the components have the same spatial distribution at all frequencies and vary only in intensity. 
This is a reasonable approximation for the diffuse foreground components we consider in this work (Galactic synchrotron and free-free emission). The spatial variation of their frequency spectra is due to changes in the properties of the inter-Galactic medium responsible for the emission (see Sec.~\ref{sec:foreg}). It is therefore reasonable to assume that these properties vary smoothly and not significantly over limited regions of the sky.

The 21\,cm signal, however, varies with frequency much more than foregrounds. In this case, a different frequency corresponds to a different redshift, and the spatial variation is due to the combined effect of changing position along the line of sight and evolution. As a result, the 21\,cm signal is expected to be essentially uncorrelated over frequency separations of the order of MHz \citep[see, e.g,][]{Bharadwaj2005,santos2005,mellema2006}. The correlation between subsequent redshift slices for our simulation is computed in Sec.~\ref{sec:21cm}.

Following \cite{chapman2012,chapman2013}, we can model the 21\,cm signal as a noise contribution. When applying eq.~(\ref{modhcca}) to our case, we interpret the source vector $\bmath{s}$ as containing the foreground components only, and we model the noise vector as $\bmath{n}=\bmath{n}_{\rm inst}+ \bmath{n}_{\rm HI}$, where $\bmath{n}_{\rm inst}$ is the instrumental noise and $\bmath{n}_{\rm HI}$ is the signal of interest.

\subsection{Foreground cleaning}
Starting from the linear mixture model of eq.~(\ref{modhcca}), we can obtain an estimate $\bmath{ \hat s}$ of the components $\bmath{s}$ through a suitable linear mixture of the data: 
\begin{equation}\label{recon}
\bmath{ \hat s}=\bmath{\sf W}\bmath{x},
\end{equation}
where $\bmath{\sf W}$ is called reconstruction matrix. In this work we use a reconstruction matrix given by
\begin{equation}\label{gls}
\bmath{\sf W}=[\bmath{\sf \hat H_{B}}^{T}\bmath{\sf C}_{\rm n}^{-1}\bmath{\sf \hat H_B}]^{-1}\bmath{\sf \hat H_B}^{T}\bmath{\sf C}_{\rm n}^{-1},
\end{equation}
which is called the generalised least square solution (GLS). It depends on the noise covariance matrix $\bmath{\sf C}_{\rm n}$ and on $\bmath{\sf \hat H_B}=\bmath{\sf B} \bmath{\sf \hat H}$, where $\bmath{\sf \hat H}$ is the estimated mixing matrix. The beam matrix is necessary when we work with frequency maps of different resolution; in this case we recover deconvolved components which we need to convolve again with the beam at each frequency. If we work with common resolution data, the beam matrix can be substituted with the Identity matrix and no deconvolution/convolution is performed.   

We finally clean the frequency maps in Fourier space by subtracting the reconstructed foreground components scaled by means of the estimated mixing matrix 
\begin{equation}\label{subtr}
\bmath{n}_{\rm HI}+\bmath{n}_{\rm inst}=\bmath{x}-\bmath{\sf \hat H}\bmath{\hat{s}}.
\end{equation}
%
As a refinement to this simple subtraction scheme, we may generalise eq.~(\ref{subtr}) as 
\begin{equation}\label{subtr2}
\bmath{n}_{\rm HI}+\bmath{n}_{\rm inst}=\bmath{x}-\bmath{\sf R} \bmath{\sf \hat H}\bmath{\hat{s}},
\end{equation}
where $\bmath{\sf R}$ is a diagonal $N_{\rm f} \times N_{\rm f}$ matrix whose diagonal elements, $r_{ii}$, are chosen to improve the subtraction. Specifically, each of them is a constant factor to adjust the amplitude of the predicted foreground contamination to be subtracted at a given frequency. Such small adjustments may be necessary as a result of errors in the foreground model. 

The estimation of the $r_{ii}$ is performed at the foreground subtraction stage, when both the mixing matrix $\bmath{\sf \hat H}$ and the foreground components $\bmath{\hat s}$ have been recovered. For each frequency channel $i$, $r_{ii}$ can be found by minimizing the power spectrum of the residual $\bmath{n}_{\rm HI}+\bmath{n}_{\rm inst}$ for the considered frequency. In the present application, we used as a figure of merit the integral of the residual power spectrum within a specified $k$ range. This is conceptually similar to minimizing the variance of the solution, but with the additional option of selecting the angular scales that are dominated by foreground emission rather than noise or 21\,cm signal. We will come back to this in Sect.~\ref{sec:three}. Since in this case the parameter space is one-dimensional and we need to sample a limited range around 1, we implemented this minimization simply as a grid search.



\section{Simulated data} \label{sec:two}
\subsection{EoR signal}\label{sec:21cm}
We used the  semi-numerical code {\tt 21cmFast} \citep{mesinger2011} to generate 3D realizations of the 21\,cm signal, the brightness temperature $T_{\rm b}$, as a function of redshift. The {\tt 21cmFast} code uses the excursion-set formalism and perturbation theory; it runs much quicker than hydrodynamic simulations and produces accurate results up to scales of $\sim 1$\,Mpc. 
We adopted the best-fit cosmological model from \cite{planck-cosmo} (\emph{Planck+WP} results), defined by $\Omega_{\rm m}=0.315$, $\Omega_{\rm b}=0.046$, $H_0=67.3$, $\sigma_8=0.829$, $n_{\rm S}=0.96$. Besides the cosmological parameters, there are other ones describing the reionization mechanism, which are poorly known and have a significant effect on the amplitude of the 21\,cm signal as a function of redshift. One of the most important is the reionization efficiency, $\zeta$ \citep{FZH2004}, which determines the mass of ionised material per unit mass of the overdensity. For our simulation we adopt  $\zeta=25$.



We simulated the evolution of the 21\,cm signal in a box of comoving size of $(1200$\,Mpc$)^3$ from $z=6$ to $z=13$ through 70 redshifts spaced by $\Delta z=0.1$. The $T_{\rm b}$ maps are slices (of thickness 2.4\,Mpc) extracted from each box after choosing one direction as the line of sight (LoS). The $\Delta z=0.1$ separation between the boxes corresponds to a comoving separation 
\begin{equation}\label{comoving}
\Delta r_i^{i+1}=\int_{z_i}^{z_{i+1}}c/H(z')dz'
\end{equation}
between two consecutive slices; such separation is larger at low redshift and smaller at high redshift (for our fiducial cosmology, $\Delta r\sim 40$\,Mpc at $z=6$ and $\sim 15$\,Mpc at $z=13$). We used eq.~(\ref{comoving}) to compute the position of the slice along the LoS for each cube. This mimics the realistic situation, in which the 21\,cm signal varies with frequency as an effect of evolution (change of the redshift box) and distance from the observer (change of the position of the slice within the box). We finally cut the 21\,cm maps to have an angular size of $3.25^\circ\times3.25^\circ$ (the common field of view for all frequencies, see next section) at each redshift, and regridded it to $256\times256$ pixels. 

\begin{figure}
\includegraphics[width=6cm,angle=90]{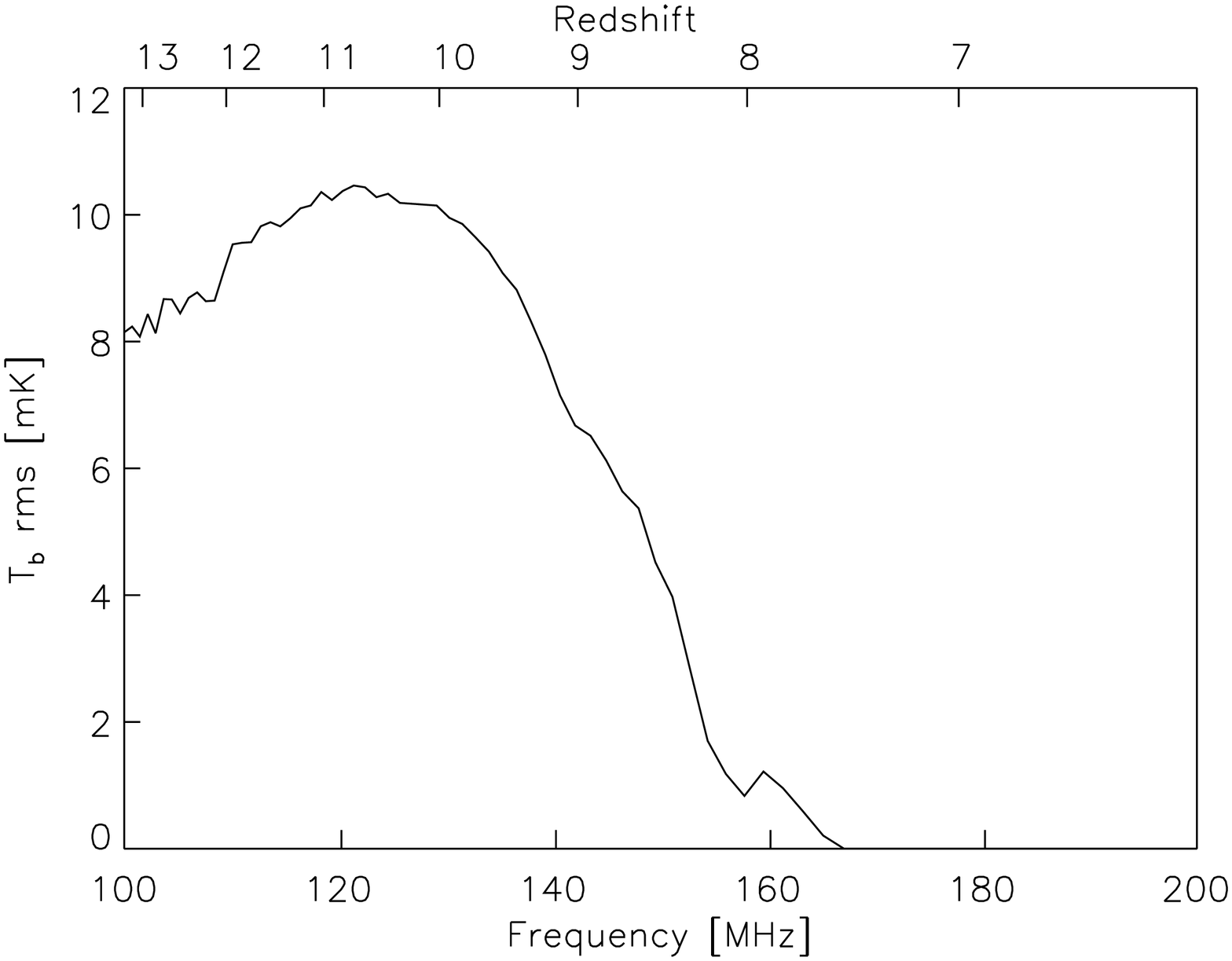}
\includegraphics[width=8cm]{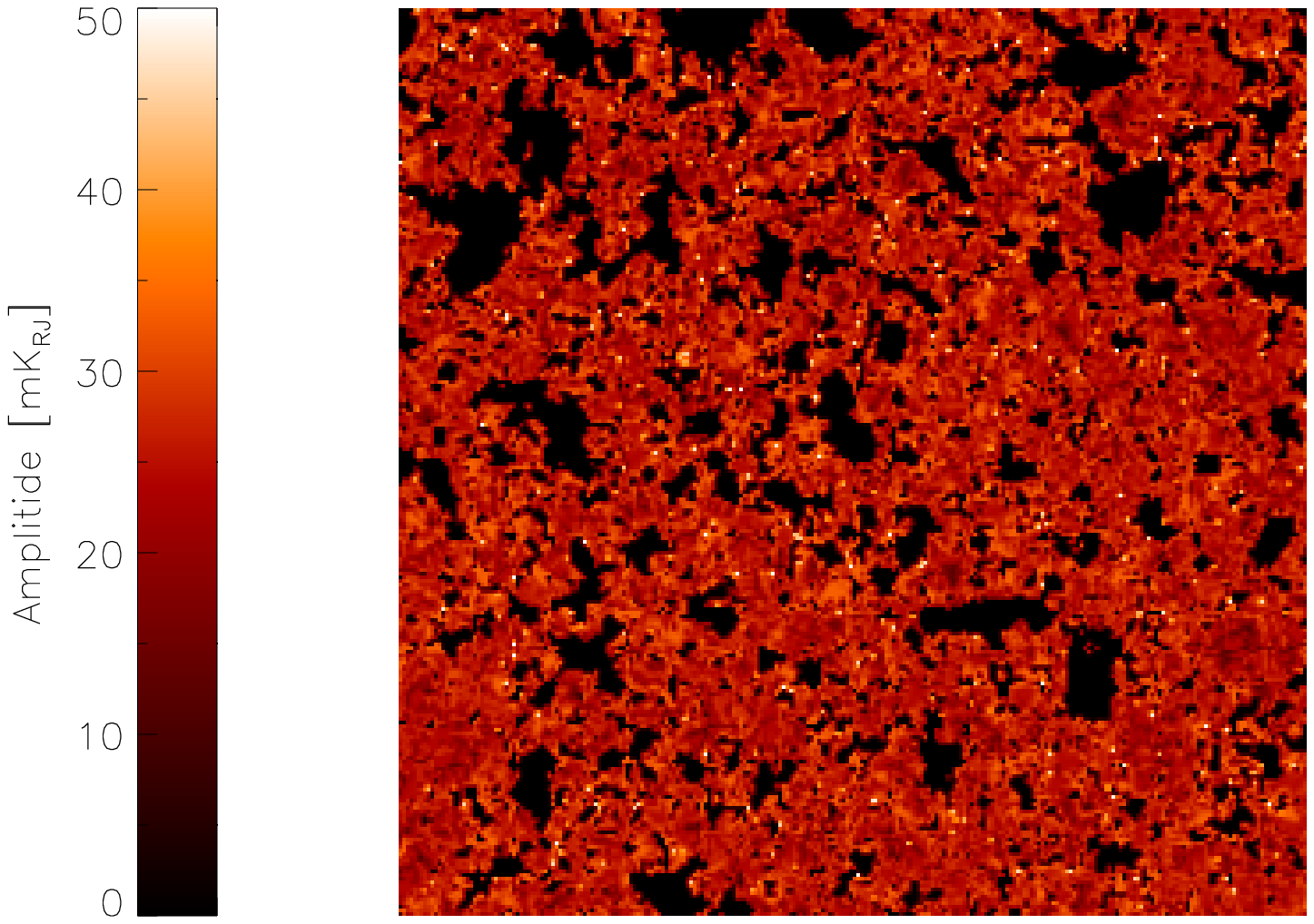}
\caption{Simulated 21\,cm signal: rms per frequency for a pixel size of $\sim 0.8$\,arcmin (top) and map at $z=11$ in units mK$_{\rm RJ}$. The sky patch is $3.25^\circ \times 3.25^\circ$ wide.}
\label{fig:21cm}
\end{figure}

In Fig.~\ref{fig:21cm} we show the rms per redshift (frequency) of the 21\,cm maps (for a pixel size $\sim 0.8$\,arcmin) and a map of the signal at $z=11$. The signal has an amplitude of $\sim 8$\,mK at $z=13$, which increases to $\sim 11$\,mK at $z=11.5$ and then decreases until it disappears around $z=7.5$. As already mentioned, this particular behaviour depends on the reionization parameters used for the simulation. 


\begin{figure}
\includegraphics[width=6cm,angle=90]{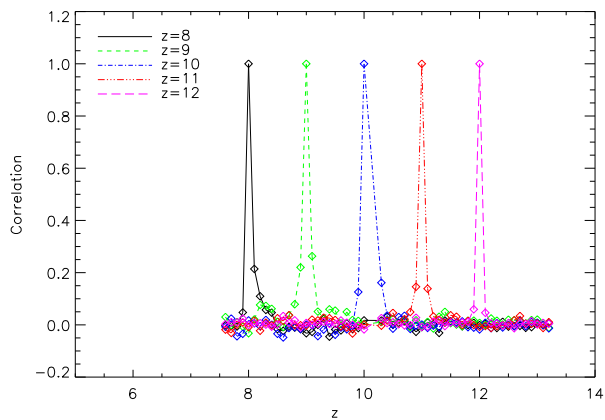}
\caption{Pearson correlation coefficients between the simulated 21\,cm maps at $z=8$, 9, 10, 11 and 12, as detailed in the legend, and all other redshifts. The maps are separated by $\Delta z=0.1$. No correlation is shown at $z<7.5$ because the signal is null.}
\label{fig:corr}
\end{figure}

In Fig.~\ref{fig:corr} we show the Pearson correlation coefficients between the 21\,cm maps of the simulation at some representative redshifts ($z=8,9,10,11$ and 12) and all others. No correlation is shown for $z<7.5$ because the 21\,cm signal is null. The measured coefficients allow us to test the hypothesis on the statistics of the signal we made in Sec.~\ref{sec:himodel}, namely that it could be modelled as a noise term, uncorrelated between frequencies. Indeed, maps separated by $\Delta z=0.1$ are still correlated at the $10$--20\% level; they can more safely assumed to be uncorrelated at $\Delta z=0.2$. Such separation is larger than what is typically considered for EoR experiments. Given the possible implications for component separation, the correlation of the HI signal between different frequency maps should not be neglected in the simulations. 

\subsection{Foregrounds}~\label{sec:foreg}
We neglect in this work the contamination from discrete sources, either resolved or as a background, and only consider the diffuse synchrotron and free-free emission from our Galaxy.

Synchrotron radiation is due to cosmic-ray electrons spiralling in the Galactic magnetic field. Its frequency behaviour reflects the spectrum of the electrons and can be described to first order by a power-law model with spectral index $\beta_{\rm s}$. At frequencies below a few GHz, $\beta_{\rm s}$ is ranging from $-2.5$ to $-2.7$ as a function of the position in the sky \citep{broad1989}. 

Free-free radiation is due to brehmstrahlung emission. Its spectrum is quite uniform and can be predicted to good accuracy; in the optically-thin regime, which holds at high latitudes, it is well approximated between 100 and 200\,MHz by a power-law with spectral index of -2.08 \citep{draine2011}.

%

%

To simulate the synchrotron and free-free emission we based on existing foreground templates: the \cite{haslam} 408\,MHz map reprocessed by \cite{remazeilles2014} \footnote{http://lambda.gsfc.nasa.gov/product/foreground/2014\_\\haslam\_408\_info.cfm} and the H$\alpha$ map described in \cite{dickinson}, respectively. Using real data as a template for the emission components preserves the non-Gaussian statistics of the foreground emission and the spatial correlation between the two components. These properties could be relevant for component separation purposes and are difficult to reproduce when simulating the components as a random realization.  

\begin{figure}
\includegraphics[width=6cm,angle=90]{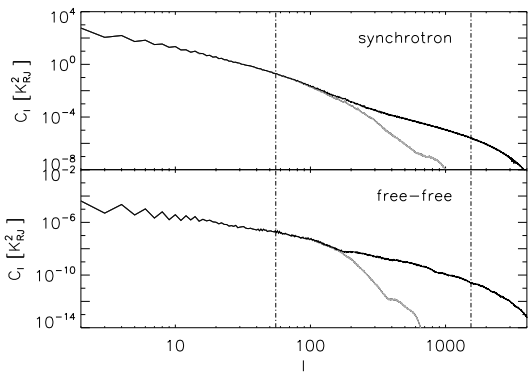}
\caption{Full-sky power spectra of the original (grey) and high-resolution (black) synchrotron and free-free templates at 408\,MHz. The vertical lines at lower and higher multipole correspond to the size of the sky patch and the size of the instrumental beam, respectively.}
\label{fig:synch_templ}
\end{figure}

We upgraded the templates in resolution from the original one ($\sim 1^\circ$) by adding to each map a Gaussian random field. We used for this field a power-law behaviour $\ell^{-\beta}$, which is consistent with what observed for these components, from $\ell=200$ ($\theta=1^\circ$) to $\ell=4000$ ($\theta=3$\,arcmins). The spectral index $\beta$ and the intensity have been chosen to obtain a smooth power spectrum of the final template around the scale of the original beam. The processing of the full-sky templates has been done with the Healpix \citep{gorski} package. In Fig.~\ref{fig:synch_templ} we show the full-sky power spectra of the original templates and the high-resolution ones. 
%

We extracted $3.25^\circ\times3.25^\circ$ patches from the high resolution templates. The data on the sphere are projected on the plane tangential to the centre of the patch and re-gridded with a suitable number of bins in order to correctly sample the original resolution. Each pixel in the projected image is associated with a specific vector normal to the tangential plane and it assumes the value of the HEALPix pixel nearest to the corresponding position on the sphere. Clearly, the projection and re-gridding process will create some distortion in the image at small scales. However, we verified that this has negligible impact on the scales considered in this work.

\begin{figure}
\includegraphics[width=8cm]{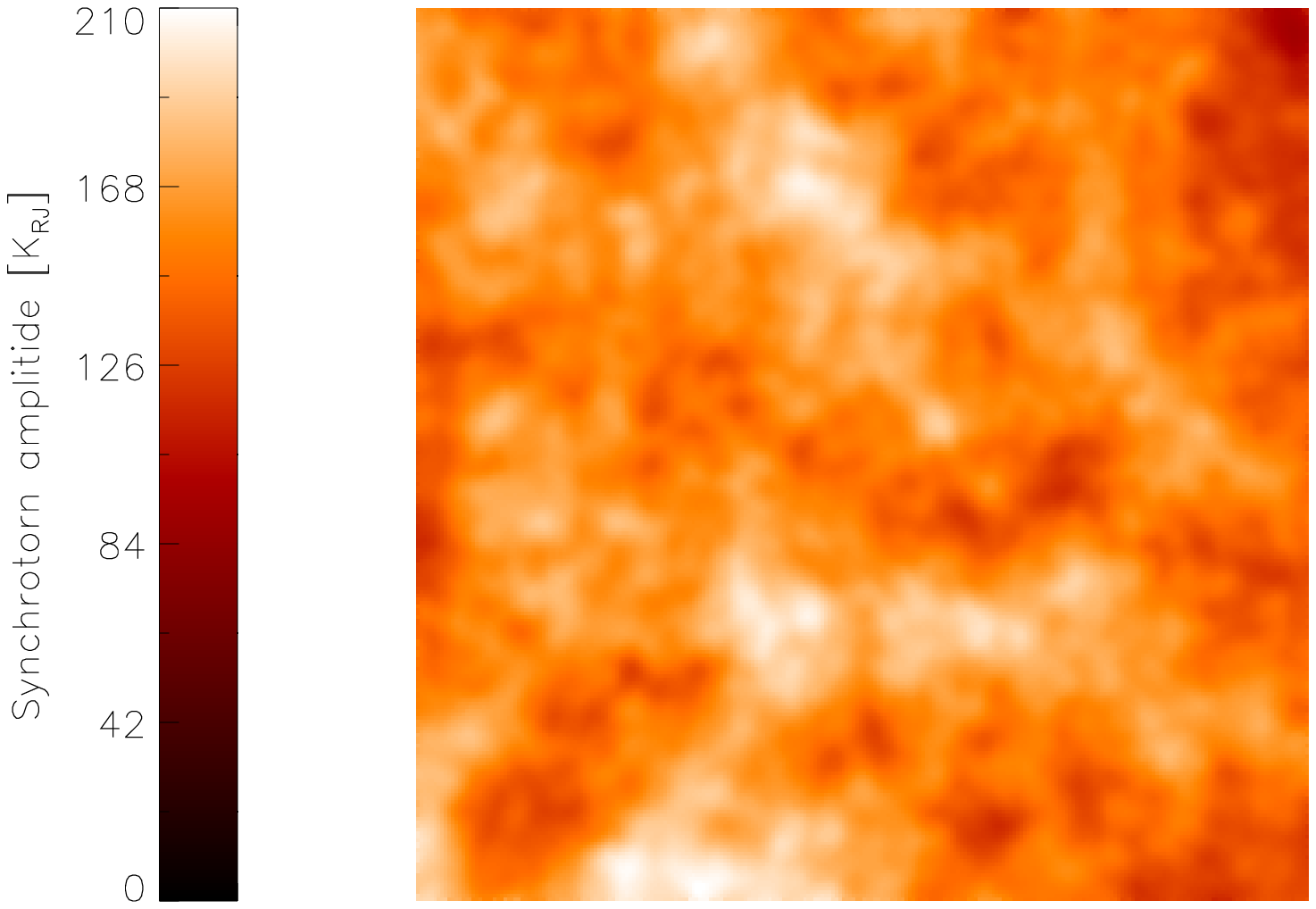}
\includegraphics[width=8cm]{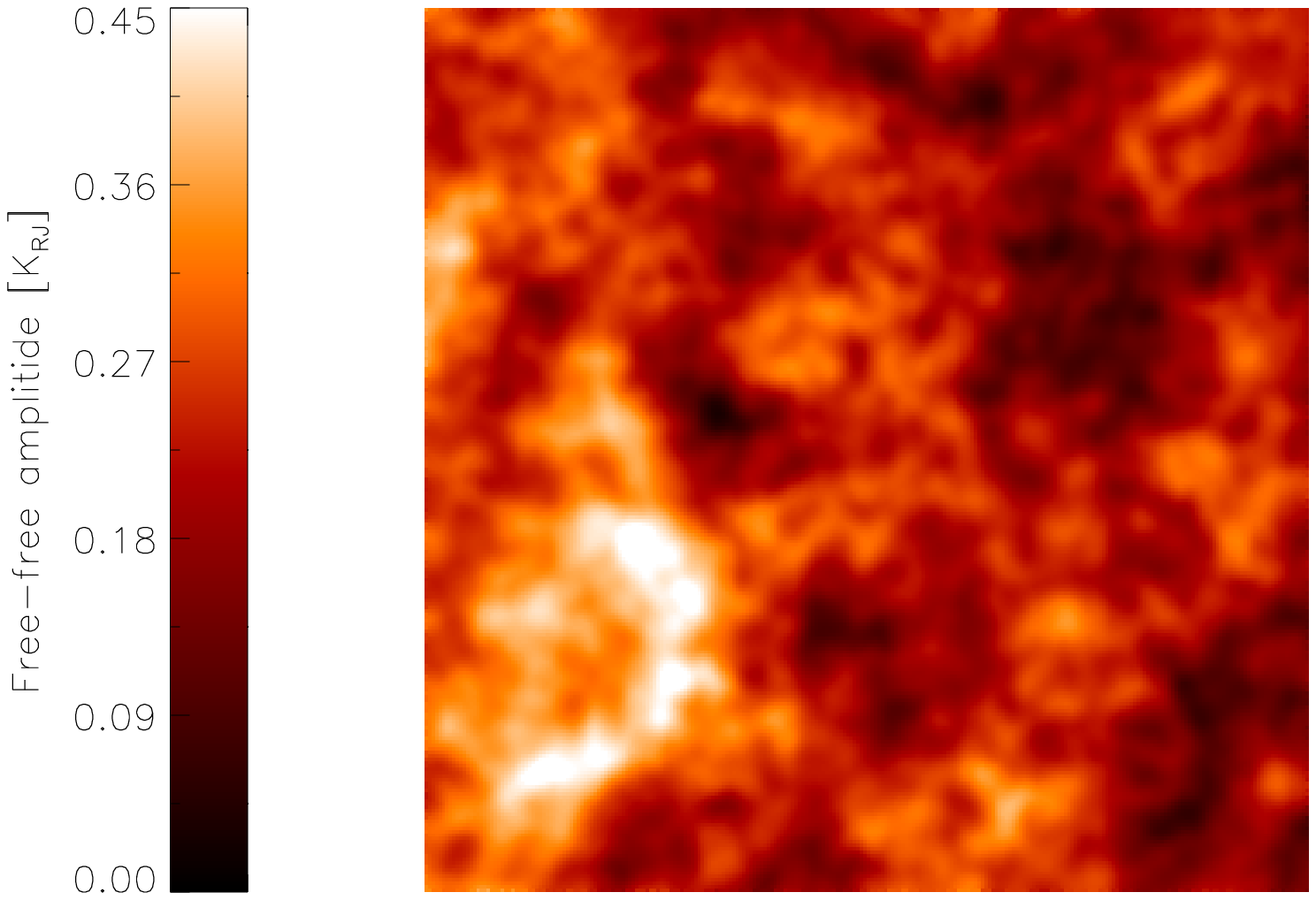}
\caption{Foreground realization at 150\,MHz in K$_{\rm RJ}$. The sky patch is $3.25^\circ \times 3.25^\circ$ wide.}
\label{fig:foreg}
\end{figure}

By choosing different central coordinates for the patches we obtain different ``realizations'' of the foreground emission components, which are realistic in terms of their auto- and cross-correlation, at least at large scales. We produced a sample of ten foreground realizations extracted from a different area of the full-sky templates. The intensity of the components has been scaled to match typical high-latitude values (20\,K at 325\,MHz for synchrotron and two orders of magnitude lower for free-free, \citealt{jelic2008}).  This means that different realizations have similar foreground contamination but a different morphology of the synchrotron and free-free amplitudes. A foreground realization at 150\,MHz is shown in Fig.~\ref{fig:foreg}.

We scaled the free-free component in frequency with a power-law model, $I(\nu)\propto \nu^{\beta_{\rm ff}}$, with $\beta_{\rm ff}=-2.08$. For the synchrotron component we adopted four simulated spectra of different complexity, which we called S0, S1, S2 and S3:
\begin{itemize}
\item S0: power-law spectrum $I(\nu)\propto \nu^{\beta_{\rm s}}$, with constant spectral index $\beta_{\rm s}=-2.6$. This model represents the ideal case, where there are no spatial variations or departure from the smooth power-law behaviour. This is not a realistic model, but it is useful as a term of comparison for more complicated models. 
\item S1: power-law spectrum $I(\nu)\propto \nu^{\beta_{\rm s}}$, with spatially-varying spectral index. The spectral index map is a random field with mean of -2.6 and standard deviation of 0.02 on the $3.25^\circ \times 3.25^\circ$ sky patch and for a pixel size of 0.8\,arcmin. The random field has a power-law behaviour in $\ell$, similar to what is expected for the amplitude of the components. 
\item S2: non-smooth spectrum oscillating around a power-law with $\beta_{\rm s}=-2.6$. The oscillations (at the 3\,\% level) are obtained by multiplying the power-law spectrum by random numbers having a Gaussian distribution with mean of 1 and standard deviation of 0.03. Rather than being physically motivated, this model has been introduced to challenge the hypothesis of spectral smoothness, which is used by many EoR component separation approaches.  
\item S3: non-parametric, curved synchrotron spectrum produced with the {\tt GalProp} code \citep{galprop}. This spectrum is physically motivated and it exploits models for the Galactic magnetic field. For this particular model, the spectral index steepens significantly in the frequency range of interest: from $\sim -2.65$ at $\nu=100$\,MHz to $\sim -2.85$ at $\nu=200$\,MHz.   
\end{itemize}

\begin{figure*}
\includegraphics[width=6cm,angle=90]{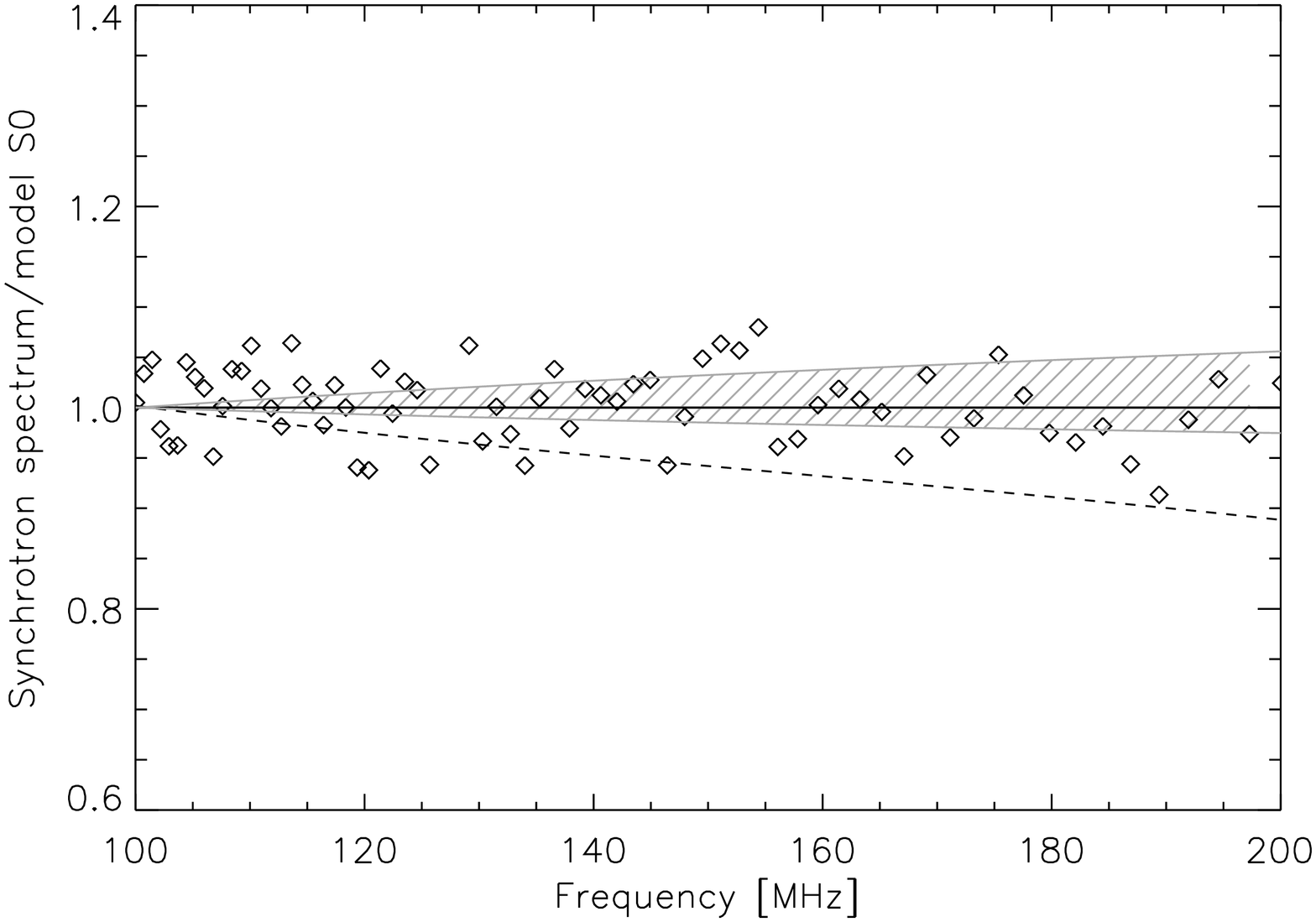}
\includegraphics[width=8cm]{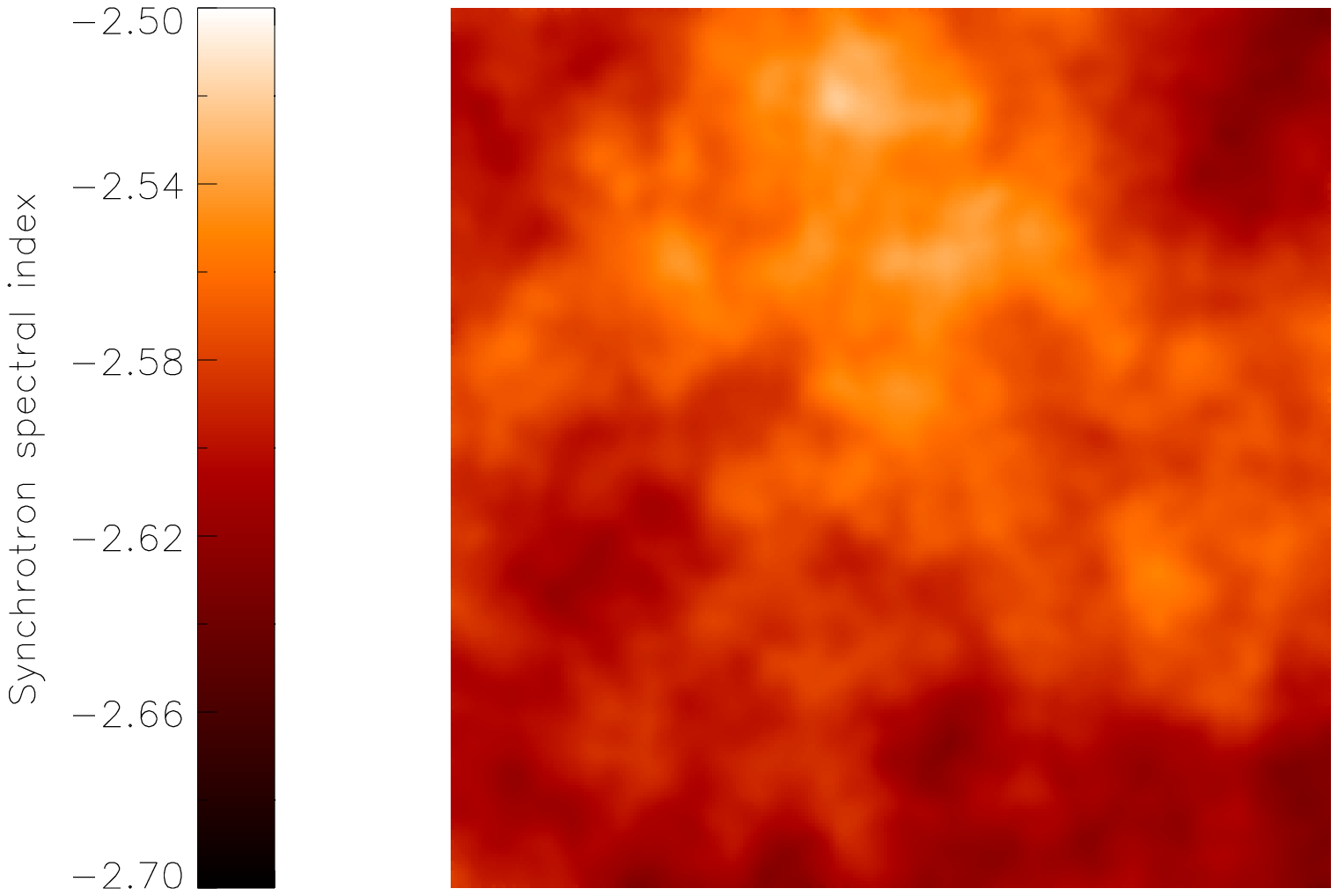}
\caption{\emph{Left}: Synchrotron spectra for all the considered models divided by the spectrum S0. Black solid line: spectrum S0; grey shaded region: area spanned by the spatially-varying spectra of S1; diamonds: spectrum S2; black dashed line: spectrum S3. All the spectra are normalised to be the same at 100\,MHz. \emph{Right}: spectral index map for the spectrum S2.}
\label{fig:foreg_models}
\end{figure*}

In the left panel of Fig.~\ref{fig:foreg_models} we compare the synchrotron spectrum for the 4 models considered; in the right panel we show the spectral index map for simulation S1 and the foreground realization of Fig.~\ref{fig:foreg}.
\subsection{Instrument} \label{sec:inst}
The SKA instrument relevant for EoR observation is the low frequency aperture array (LFAA). This is an aperture array consisting of around 260 thousands wide-bandwidth antennas of a single design, to be mounted on the Australian SKA site. The station diameter is around 35\,m. The configuration is very compact, with 75\% of the antennas within a 2\,km diameter core. The frequency coverage is from 50 to 350\,MHz. 

In this work we consider only the core of the low-frequency array for SKA phase 1 over a frequency range of 100-200\,MHz. With these instrumental specifications we obtain a field of view (FoV) of $\sim6.5^\circ\times6.5^\circ$ and a synthesized beam of FWHM $\sim6.7$\,arcmins at 100\,MHz, which become  $\sim3.25^\circ\times3.25^\circ$ and $\sim3.35$\,arcmins respectively at 200\,MHz.
However, building a simulation with frequency-dependent beam is quite computationally demanding. For this work, we construct a mask for the $uv$ plane that only contains the $uv$ points sampled at all frequencies. By applying this mask to the data in the Fourier domain we equalize the resolution of the maps between frequencies. The price we pay is a data loss, which results in a worsening of the instrumental performances. 



\begin{figure}
\includegraphics[width=6cm,angle=90]{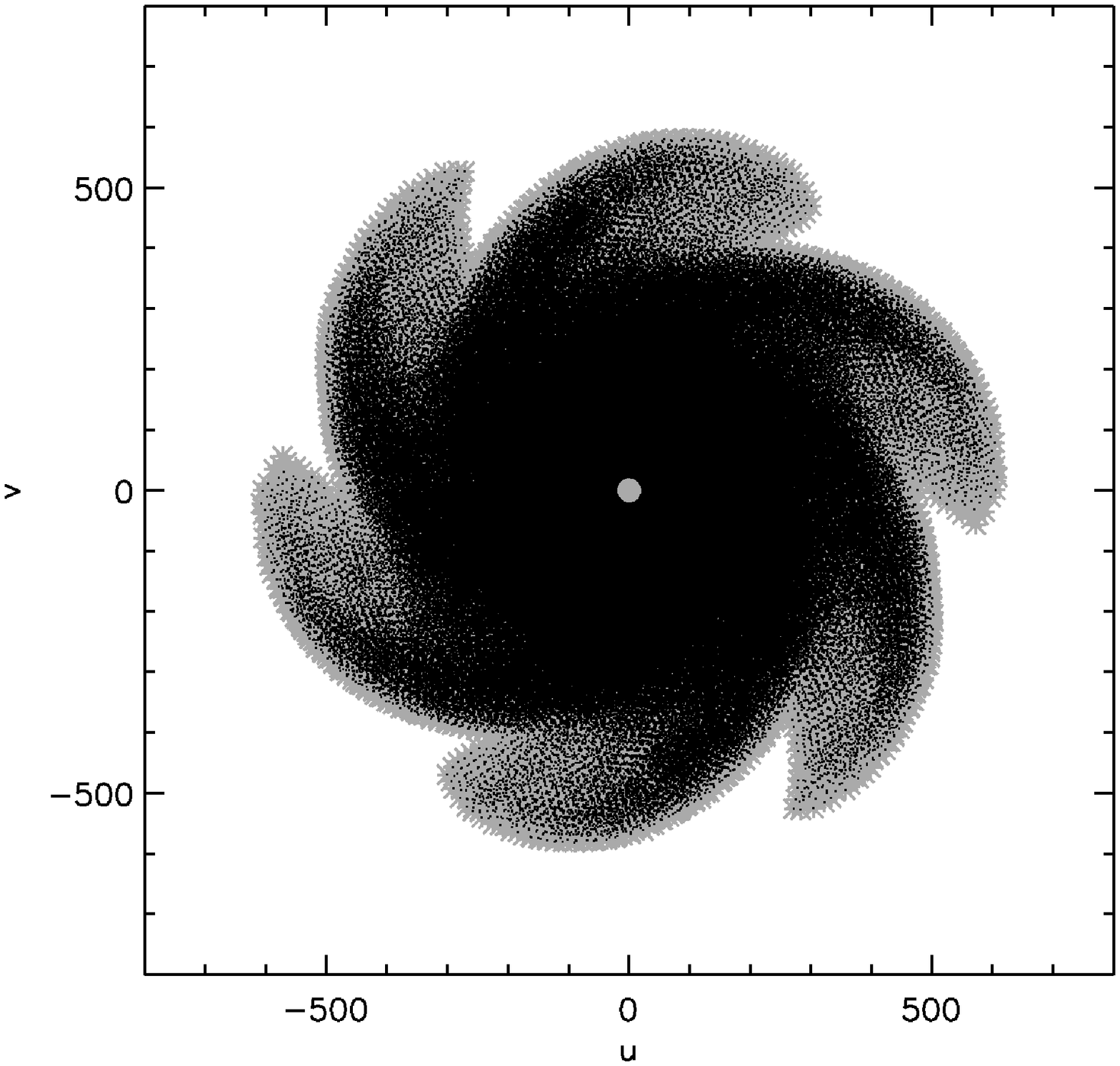}
\includegraphics[width=6cm,angle=90]{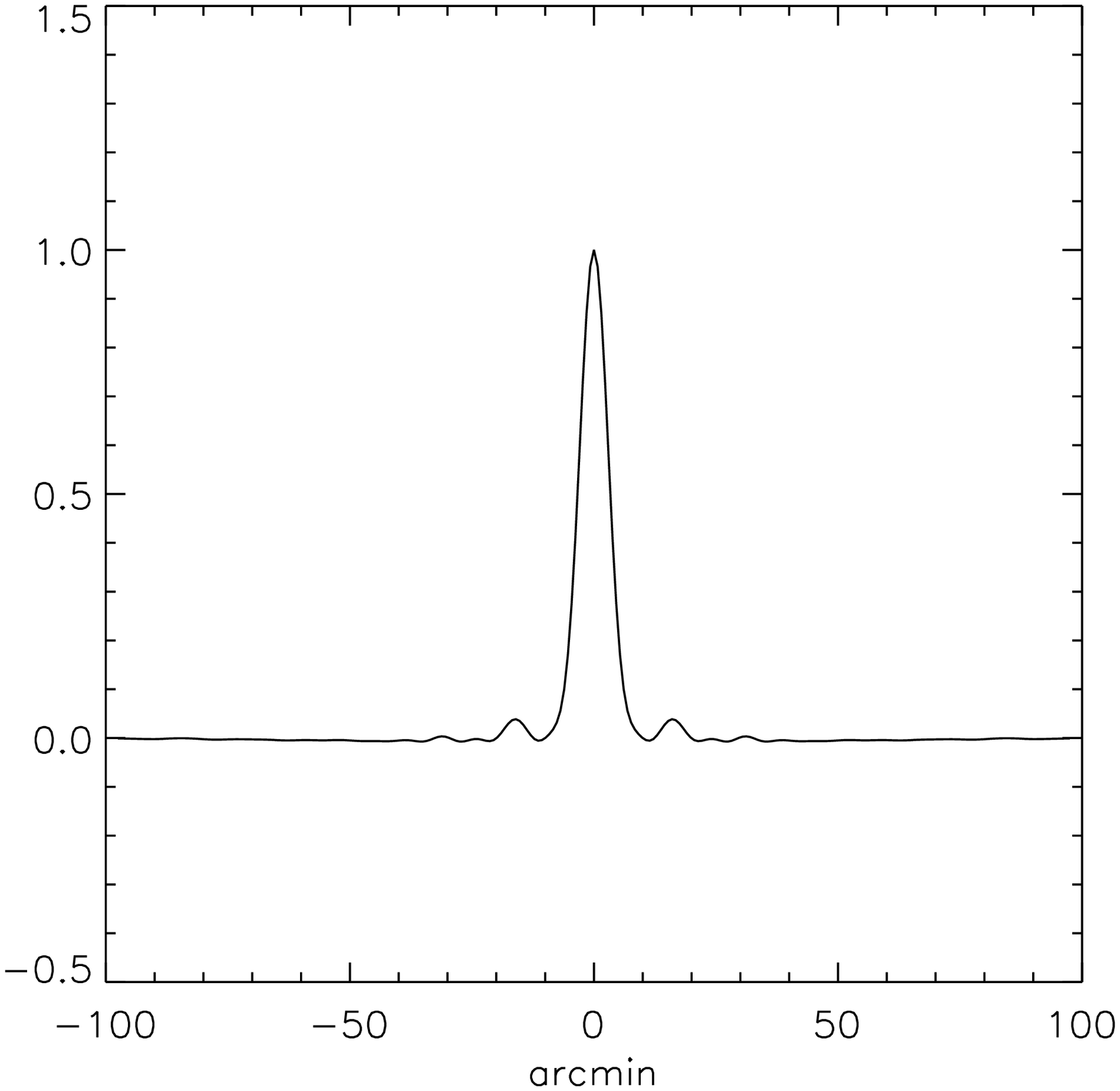}
\caption{Top: $uv$ coverage containing the points sampled at all frequencies (black dots), for a Zenith observation and neglecting the rotation of the sky. The grey crosses show the additional points sampled at 100\,MHz that are lost after the $uv$ masking. Bottom: cut of the dirty beam corresponding to the coverage shown in the top panel.} 
\label{fig:inst}
\end{figure}

In Fig.~\ref{fig:inst} we show the $uv$ coverage we used for all frequencies, obtained by selecting the points sampled at all bands from 100 to 200\,MHz.  The single-frequency coverage corresponds to a proposed array configuration and a Zenith observation; the rotation of the sky has been neglected for simplicity.
The masking of the $uv$ plane results in a loss of $\sim 3$\,\% of the $uv$ samples with respect to a single frequency observation (compare the black points to the grey ones in Fig.~\ref{fig:inst}, corresponding to 100\,MHz). A cut of the dirty beam corresponding to this sampling is show in the bottom panel of Fig~\ref{fig:inst}. The synthesized beam has a FWHM of 6.7\,arcmin and the common FoV is $3.25^\circ \times 3.25^\circ$. We stress that these specifications are pessimistic: by considering the whole array instead of just the core, and by optimising the way we deal with the frequency-dependent beam and FoV, these can be improved significantly.

The sensitivity requirement for SKA phase 1 to measure the EoR signal is an rms noise of $\sim 1$\,mK on scales of 5\,arcmins at frequencies around 100\,MHz. For a given array configuration, the exact noise levels depend on the integration time, the declination of the source and the bandwidth. They also depend on frequency, since both the total noise and the gain corresponding to each element of the array are a function of $\nu$. An accurate simulation of the noise properties in our case should also account for the masking in the $uv$ space that we performed to equalize the resolution. We note, however, that such a simulation is beyond the scope of this paper. The focus of this work is on foreground emission which, for a sensitive instrument such as the SKA, is orders of magnitude stronger than noise. Therefore, we simplified the noise description by neglecting the frequency dependence and we assumed a brightness sensitivity of 1\,mK over the whole frequency range. 

The noise maps have been simulated as a Gaussian random field in the $uv$ space and filtered with the $uv$ coverage shown in Fig.~\ref{fig:inst}. They have been subsequently transformed to the pixel space and divided by the rms per pixel (computed for a pixel size of $3.04 \times 3.04$\,arcmins) to obtain the 1\,mK rms level.



\section{Analysis}\label{sec:analysis}
\subsection{Fitting for the synchrotron spectrum}\label{sec:cca_fit}
We fitted for the spectra of the foregrounds for the simulations S0, S1, S2 and S3 with the CCA method described in Sect. \ref{sec:method}. The estimation has been performed for ten foreground realizations (see Sect. \ref{sec:foreg}). We modelled the mixing matrix as consisting of  two components, free-free and synchrotron. We assumed the free-free spectrum to be known and focussed on the estimation of the synchrotron spectrum. 
This choice is reasonable because the uncertainties on the spectrum of the free-free emission are much smaller than those on the synchrotron emission. Moreover, the free-free emission is orders of magnitude fainter than synchrotron, so that the uncertainty on the latter component dominates the error budget.


In principle, we should characterize the noise component as consisting of both instrumental noise and HI signal. This is particularly true for the SKA, given that the instrumental noise is lower than the expected signal for a wide redshift range. Nonetheless, we characterised only the instrumental noise, not to exploit our previous knowledge of the intensity of the simulated 21\,cm signal. We verified that, because the foreground emission is so bright, the estimated synchrotron spectrum is very stable against sensible changes of the assumed noise levels.

We modelled the synchrotron emission as a power law, and fitted for a synchrotron spectral index. This model is fully adequate only for S0 and it is simpler than the actual input foreground models for the other simulations. This reproduces the typical situation in which the models that we use are only an approximate representation of the real data.
In order to assess the goodness of the CCA results, for each model we computed a true ``effective'' spectral index $\beta_{\rm s}$ to be compared with the estimated one, $\hat \beta_{\rm s}$. Such effective spectral index is the true spectral index for S0; the average of the spatially-varying spectral index map for S1; the slope of the underlying smooth power-law spectrum for S2; and the slope between 100 and 200\,MHz for S3.

\begin{figure}
\begin{center}
\includegraphics[width=7cm,angle=90]{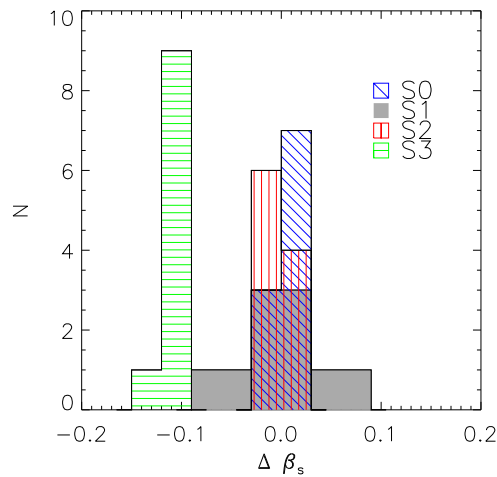}
\caption{Histogram of $\Delta \beta_{\rm s}= \hat{\beta}_{\rm s}-\beta_{\rm s}$ (estimated index$-$true effective index) for 10 foreground realizations and for the 4 simulated synchrotron spectra as detailed in the legend.} 
\label{fig:histog}
\end{center}
\end{figure}

In Fig.~\ref{fig:histog} we show the histogram of the errors $\Delta \beta_{\rm s}=\hat{\beta}_{\rm s}-\beta_{\rm s}$ for the 10 foreground realizations and the 4 spectral models considered. This represents the random estimation error due to the CCA method, without considering errors due to incorrect modelling of the true synchrotron spectrum. Given the strength of the contaminants, it is very important that the random error is small.

For S0 and S2 the estimation is very good both in terms of width of the distribution and offset from zero. 
Spatial variations of the spectral index (S1) mainly broaden the error distribution while keeping the offset from zero low. Conversely, the steepening of the spectrum (S3) mainly shifts the distribution from zero without enlarging it. In this case, the estimated index is a good representation of the spectral index at low frequencies, where the synchrotron emission is stronger and the spectral index is flatter. 

Overall, the CCA method is able to fit the slope of the spectrum with good accuracy, not being challenged too much by the higher complexity of the true spectrum with respect to the adopted model. For the simulation with steepening of the spectral index there is a bias towards the flattest slope. 

\subsection{Cleaned maps} \label{sec:three}
\begin{figure}
\includegraphics[width=8cm]{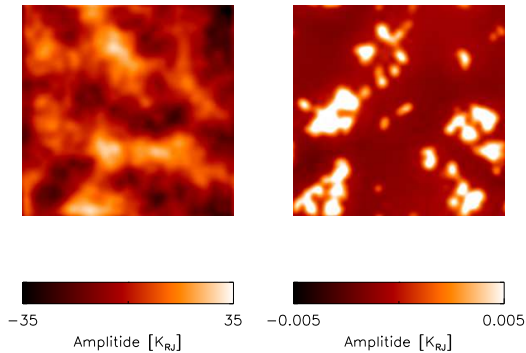}
\caption{True foreground emission (left) and HI signal (right) at 150\,MHz ($z=8.5$).} 
\label{fig:true_150}
\end{figure}

In this section we visualise the residual foreground contamination due to the random and model error on the estimated mixing matrix. We show the cleaned maps in pixel space at the central frequency ($\nu=150$\,MHz, $z=8.5$) and for the foreground realization shown in Fig.\ref{fig:foreg}. The cleaned maps are obtained by Fourier-transforming the data cleaned in the visibility plane and are convolved with the instrumental beam. 
In Fig.~\ref{fig:true_150} we show the true foreground contamination and the true HI signal. Both maps are convolved with the beam.

\subsubsection{Simulation S0}
\begin{figure}
\includegraphics[width=8cm]{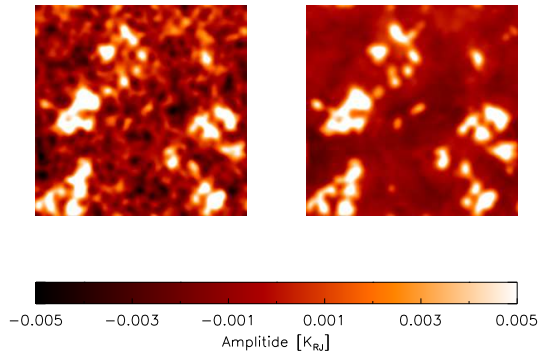}
\caption{Ideal case (S0): reconstructed HI map at 150\,MHz ($z=8.5$) with (left) and without (right) noise in the data, to be compared with the true input signal (right panel of Fig~\ref{fig:true_150}).} 
\label{fig:m0}
\end{figure}

In Fig.~\ref{fig:m0} we show the reconstructions at 150\,MHz for S0 with and without noise in the data. The residual foreground contamination is well below the noise, due to the lack of any model error and any non-idealities of the data. 
\subsubsection{Simulation S1}
The spatial variability of the synchrotron spectral index (tested by S1) is something that we cannot model explicitly in the CCA method. In fact, its second-order statistics constraint assumes that the mixing matrix is constant in the considered area of the sky. In principle, one could divide the FoV into smaller areas and process them separately. This strategy is successfully used in \cite{ricciardi2010} for CMB data, as we consider large sky areas where the spectral properties of foregrounds are likely to vary significantly. However, the sky patches cannot be arbitrarily small, because a robust computation of the data spectra and cross-spectra needs good statistics. This means that in the present application, with sky areas considered are already relatively small, this approach could be inefficient.


Alternatively, following \cite{stolyarov2005}, a single component having spatially-varying frequency scaling can be modelled as the sum of multiple components with uniform frequency scaling. A suitable decomposition is obtained by expanding the spectral model of the synchrotron component in a Taylor series around the mean spectral index. Thus, the first-order component has spectrum $I(\nu) \propto \nu^{\hat{\beta}_{\rm s}}$, the second-order one has spectrum   $I(\nu) \propto \nu^{\hat{\beta}_{\rm s}-1}$, and so on. 

\begin{figure}
\includegraphics[width=8cm]{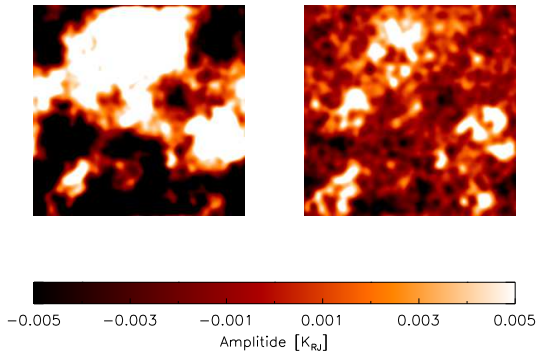}
\caption{Effect of spatially-varying synchrotron spectral index (S1): reconstructed HI map at 150\,MHz ($z=8.5$) with two (left) and three (right) effective foreground components, to be compared with the true input signal (right panel of Fig~\ref{fig:true_150}).} 
\label{fig:m1}
\end{figure}

In Fig.~\ref{fig:m1} we show the results at 150\,MHz for two reconstructions: one where we fit for the synchrotron and free-free components and one where we included an additional foreground component, with spectrum $I(\nu) \propto \nu^{\hat{\beta}_{\rm s}-1}$, to absorb the errors due to the spatial variations of the synchrotron spectral index. In the first case the errors in the foreground models cause a residual contamination which is too large compared to the signal we aim to recover. The morphology of the residual map reflects that of the spatially-varying spectral index map (shown in the right panel of Fig.~\ref{fig:foreg_models}). 

The inclusion of the extra component improves the results substantially; the residual map, reduced by $\sim 80\,\%$, is now below the 21\,cm signal. In principle, one could add more terms of the Taylor expansion to reach even higher accuracy, however the noise in the reconstruction increases with the number of components. In our case, the inclusion of a 4th component to the model does not improve appreciably the cleaning. 

\subsubsection{Simulation S2}
\begin{figure}

\includegraphics[width=8cm]{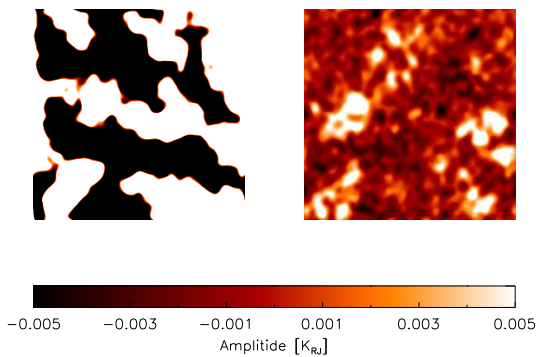}
\caption{Effect of 3\,\% random oscillations around the smooth spectrum (S2): reconstructed HI map at 150\,MHz ($z=8.5$) with simple (left) and improved (right) subtraction method, to be compared with the true input signal (right panel of Fig~\ref{fig:true_150}).}
\label{fig:m2}
\end{figure}

In the left panel of Fig.~\ref{fig:m2} we show the reconstructed HI emission at 150\,MHz for S2 obtained with the simple subtraction method [eq.~(\ref{subtr})]. The 3\,\% oscillations around the power-law spectrum are enough to swamp the detection of the HI signal. The morphology of the reconstructed map is that of the foreground component shown in the left panel of Fig.~\ref{fig:true_150}. 

In the right panel of Fig.~\ref{fig:m2} we show the results of the improved subtraction method  [eq.~(\ref{subtr2})]. This method is able to remove efficiently the foreground contamination.  
As mentioned in Sect.~\ref{sec:method}, the improved subtraction consists in estimating a factor, $r_{ii}$, to calibrate the foreground map to be subtracted from the $i$-th frequency map. These factors have been chosen as those minimising the power of the fluctuations of the residual map. Because the estimated foreground component is obtained by linearly combining all the frequency maps, it should not correlate significantly with any single HI map or noise map. Therefore, this procedure is unlikely to subtract the signal of interest. However, we minimised only the large-scale power (larger than several arcminutes) which, in the presence of subtraction errors, is dominated by foreground emission. 

\begin{figure}
\includegraphics[width=6cm,angle=90]{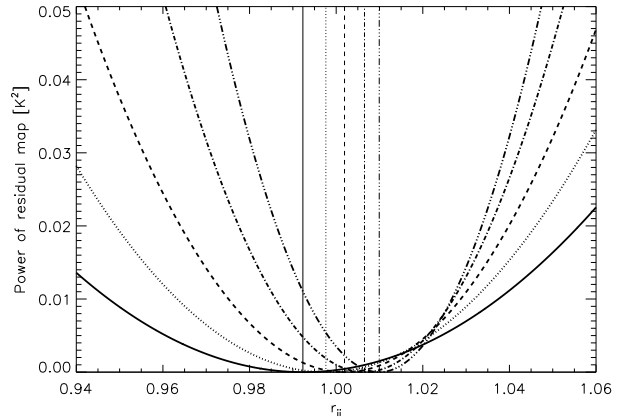}
\caption{Power of the residual map vs factors $r_{ii}$ multiplied to the predicted foreground contamination before subtraction (thick lines) and resulting optimal $r_{ii}$ value (thin lines), at five different frequencies (different line styles).}
\label{fig:recalib}
\end{figure}

As we can see from Fig.~\ref{fig:recalib}, the power of the difference map varies substantially for small variations of $r_{ii}$, and may prefer a value of $r_{ii} \neq 1$. This means that the original model either slightly underestimated or overestimated the foreground contamination, as a result of the unmodelled random fluctuations. We verified that the improved subtraction method is only necessary in the presence of significant model errors; for the random oscillations considered here, the simple subtraction method of eq.~(\ref{subtr}) is good enough if the oscillations are below 1\,\%.

%

\subsubsection{Simulation S3}
\begin{figure}
\includegraphics[width=8cm]{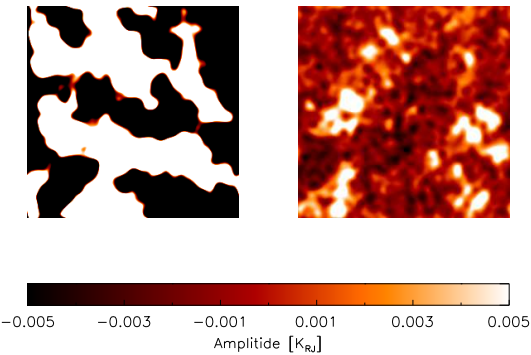}
\caption{Effect of curvature of the synchrotron spectral index (S3): reconstructed HI map at 150\,MHz ($z=8.5$) with simple (left) and improved (right) subtraction method, to be compared with the true input signal (right panel of Fig~\ref{fig:true_150}).} 
\label{fig:m3}
\end{figure}

In Fig.~\ref{fig:m3} we show the results at 150\,MHz for the simulation S3. Similarly to the previous results, the simple subtraction of the predicted power-law component (left panel) is not accurate enough, but the improved subtraction scheme of eq.~(\ref{subtr2}) (right panel) is able to correct for departures from the spectral model. 

\subsubsection{Summary}
We were able to obtain good reconstructions of the underlying HI signal for all the simulations considered. The fitted CCA spectrum achieved a very good first-order subtraction of the foreground contamination (the contamination is reduced by more than 3 orders of magnitude). Second-order corrections were necessary to compensate for departures of the true spectra from the adopted models. In particular, the simulation S1 required the addition of a further foreground component, while simulations S2 and S3 required adjustments in the subtraction method. 

We note that these two corrections are conceptually different. The addition of extra components modifies the morphology of the total foreground emission with frequency, and therefore it compensates for errors on the forground pattern, such as the one induced by a spatial variation of the spectral properties. On the other hand, the calibration of the subtraction modifies the intensity of the foreground emission and not its pattern, so it is useful in the presence of an error in the average foreground frequency spectrum, such as the one induced by an incorrect modelling. 

In a real situation, both errors are likely to be present, and these two corrections can be applied sequentially. In the next section we show results obtained by performing both corrections for all the models, irrespective to whether they are needed by the data or not. In the latter case, the amplitude of the correction results to be null or negligible. 

\section{Results}\label{sec:results}
In this Section we present a more quantitative assessment of the results for the whole frequency range considered and all the foreground models.  
\subsection{Statistics on pixel maps}
We first considered statistics on the 21\,cm maps in pixel space, and in particular the rms of the foreground-cleaned signal compared to the true one, and the Pearson's correlation coefficient between the cleaned and true signals. 
Both the cleaned and the true signals are convolved with the beam and sampled with a 0.8\,arcmin pixel size. The clean signal is noisy while the true one is noiseless.

\begin{figure}
\includegraphics[width=6cm,angle=90]{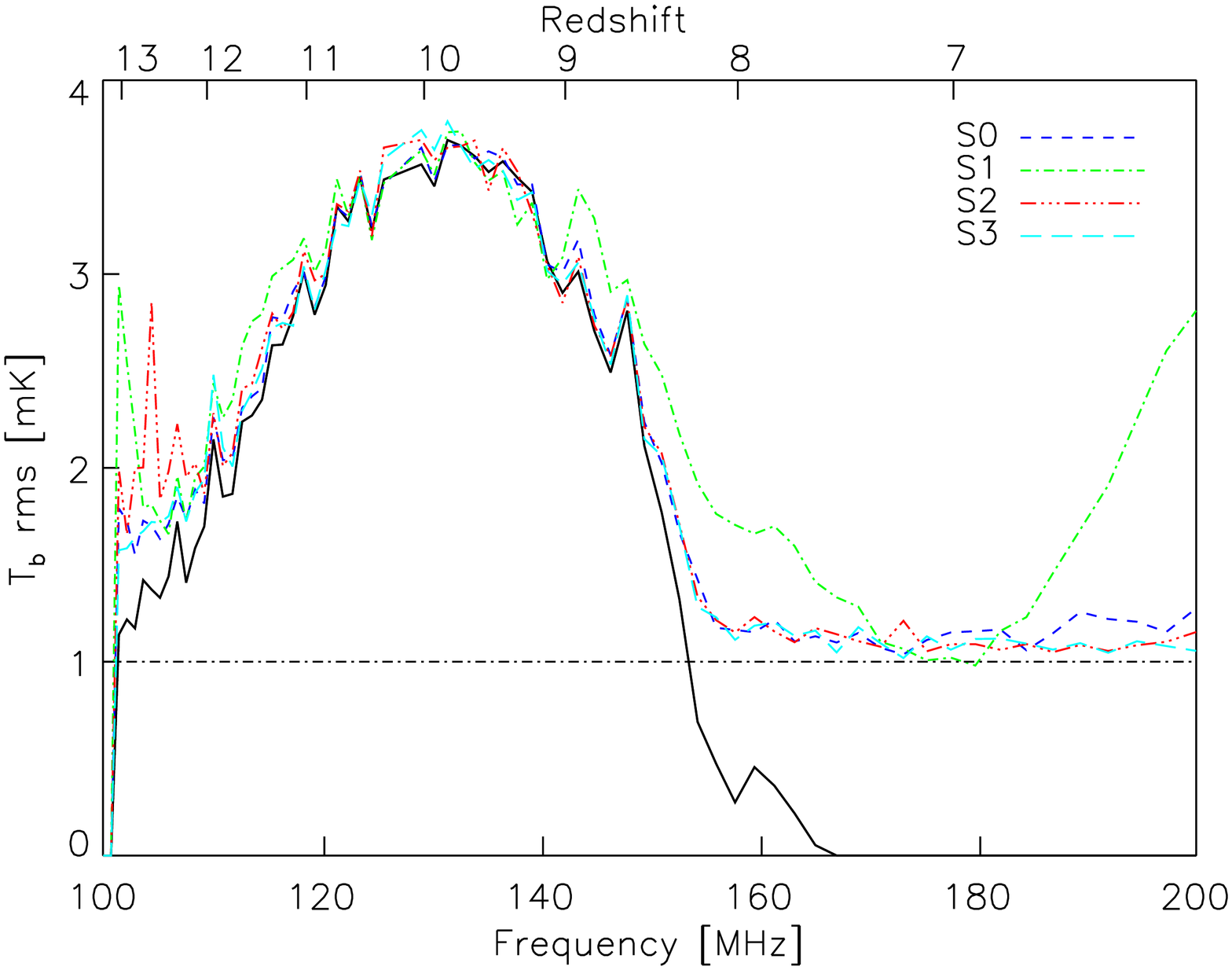}
\includegraphics[width=6cm,angle=90]{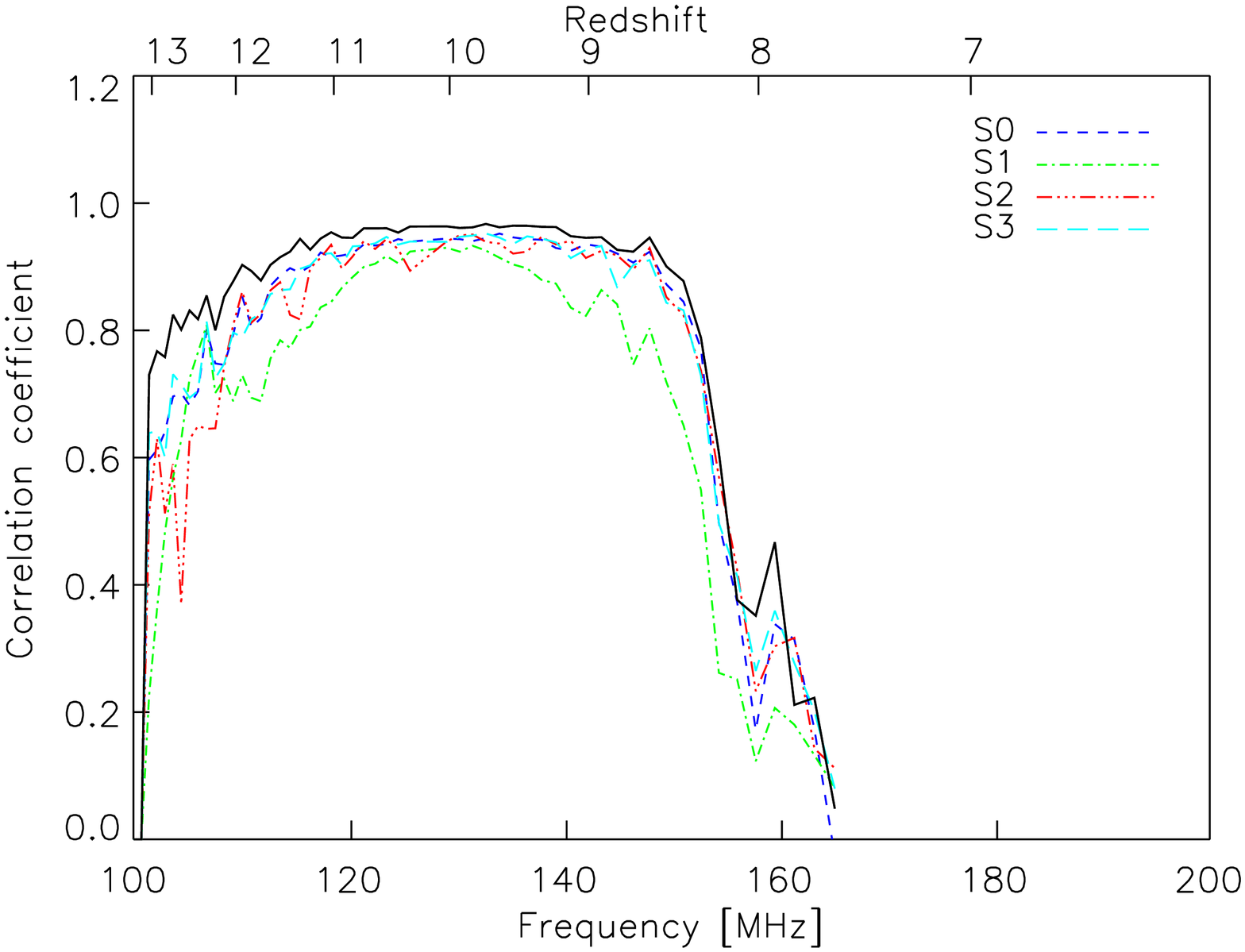}
\caption{\emph{Top}: rms of the true signal (black line) compared to rms of the foreground-subtracted signals for different models (coloured lines). The foreground-subtracted signal is noisy; the noise level is shown by the black dot-dashed line.  \emph{Bottom}: Pearson correlation coefficients between the true smoothed HI signal and the reconstructed for different models (coloured lines). The black line is the correlation with the true noisy signal and represents the best result we can achieve. Both the rms and the correlations refer to a pixel size of 0.8\,arcmin and a beam of 6.7 \,arcmin.}
\label{fig:correlations}
\end{figure}

The rms and correlation are shown respectively in the top and in the bottom panel of Fig.~\ref{fig:correlations}. Note the difference between the rms of the true signal in Figs.~\ref{fig:correlations} and 1. The signal in Fig.~\ref{fig:correlations} is significantly lower, especially at high redshift. This is the effect of the convolution of the maps with the 6.7\,arcmin instrumental beam. Since most of the power of the high-redshift signal comes from small scales, the convolution suppresses the rms significantly in this case. 

Both the rms and the correlation plots are also affected by the presence of noise in the recovered signal. To interpret the noise contribution correctly, in the rms plot we show the noise level, and in the correlation plot we show the correlation between the true noisy signal and the true noiseless signal, which represents the highest correlation we can obtain in the presence of noise.

The rms and the correlation plots give very similar indications: there is a wide frequency range $110$--150\,MHz ($z=12$--8.5) where the cleaning is very good; at higher frequencies the signal is dominated by noise while at lower one there is an excess due to residual foreground contamination. 

The details of the performance vary for the different foreground simulations considered. Overall, the worse results are obtained for S1, featuring spatial variability of the spectral properties. The unsmooth features, tested with the simulation S2, are troublesome mostly at frequency lower than 110\,MHz. It is interesting to note that the curved synchrotron spectrum performs nearly at the level of the ideal model S0. 

The pixel statistics discussed so far are easy to interpret and very useful to visualize the results. However, they may not give a complete picture because they depend on the choice of the pixel size, which is quite arbitrary, and they typically probe the signal at the smallest scales. In the next subsection we also consider 2D power spectra, which give a description of the signal, noise and residual foreground as a function of the spatial scale. 

\subsection{Power spectra} \label{sec:four}
The 2D power spectra of the 21\,cm maps, ${C}_{\rm 2D}(k)$, are computed by averaging the 2D Fourier transform of the map in circular bins corresponding to a Fourier mode $k$ [eq.~(\ref{dataspectrum})]. The results are presented in terms of $\Delta^2_{\rm 2D} (k)=\frac{A}{2\pi} {C}_{2D}(k)$, where $A$ is the area of the simulation map. 

\begin{figure}
\includegraphics[angle=90,width=8cm]{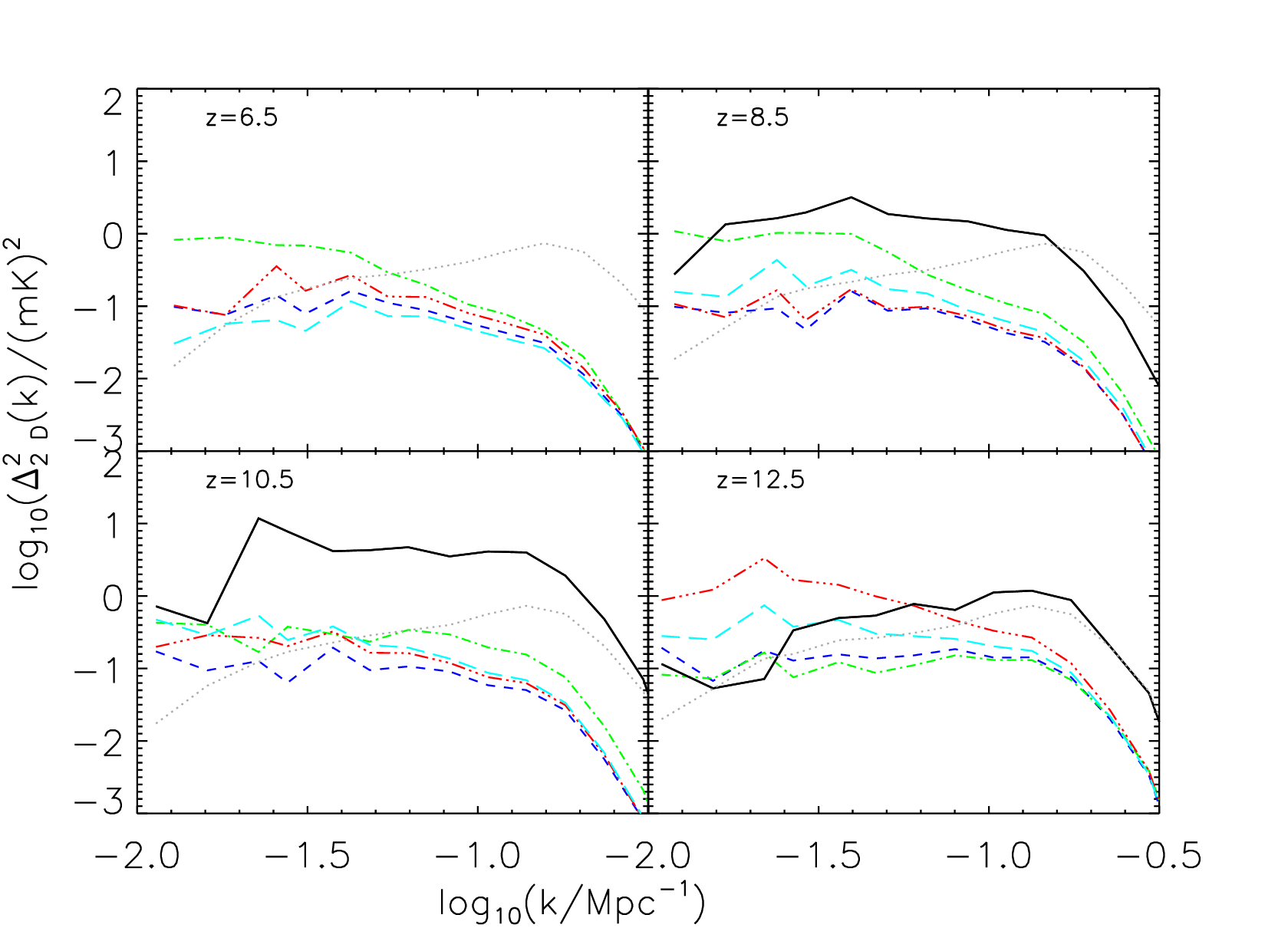}
\caption{Power spectra of the true HI signal (solid thick lines) for different redshifts, as specified in each panel, compared to the power spectrum of foreground residuals (coloured lines, colours and line styles as in Fig.\ref{fig:correlations}) and of the noise averaged for the 100 noise realizations (grey dotted lines).} 
\label{fig:2dspecs}
\end{figure}

In Fig.~\ref{fig:2dspecs} we show the power spectrum of the true convolved signal at four redshifts ($z=6.5, 8.5, 10.5$ and 12.5, thick black lines) compared to the power spectrum of the residual foreground contamination (cleaned noiseless signal $-$ true signal) for all the models (coloured lines) and the power spectrum of the noise averaged over 100 noise Monte Carlo (MC) realizations (grey dotted lines). 

The accuracy of the spectra depends on the balance between the residual foreground contamination, the noise level, and the intensity of the 21\,cm signal. At $z=6.5$, where our simulation has null HI signal, foreground residuals constitute an upper limit to the detection. At $z= 8.5$ and $z= 12.5$ the signal is above the residual contamination for a wide range of scales; the recovery of the largest scales is hampered by foreground contamination, at a level that varies for different foreground models. At $z=10.5$ we have the best recovery; foreground residuals are around two orders of magnitude below the input power spectrum. 

The simulation S1 gives the highest large-scales residuals at low redshift (high frequency) but it performs reasonably well at high redshift (low frequency). This means that, for this model, the cleaning is more efficient for the most contaminated channels. This is a consequence of the spatial variations of the synchrotron spectral index, which causes the morphology of the synchrotron component to change with frequency. Because this component is stronger at lower frequency, the reconstructed synchrotron map is more similar to the low-frequency synchrotron emission, hence it performs better when subtracted from the low-frequency channels.   

Conversely, the simulation S2 performs better at high frequency than at low frequency. The presence of non-smooth features in the spectrum causes a random error in the estimation of the mixing matrix. The resulting foreground contamination is worse where the foregrounds are stronger, i.e. at low frequency. 

For all the considered foreground models, we have a good cleaning of the signal for a wide range in scales and redshifts. This shows that our approach is a powerful one for cleaning the HI signal from foreground contamination.

\section{Conclusions}\label{sec:conclu}
We have tested the performance of the CCA component separation method \citep{bonaldi2006,ricciardi2010}, developed for CMB data analysis, on simulated SKA data to study the EoR. Our simulation includes the EoR signal, generated with  the {\tt 21cmFast} code \citep{mesinger2011}, and diffuse synchrotron and free-free foregrounds, simulated starting from the 408\,MHz \cite{haslam} map reprocessed by Remazeilles et al. (in prep) and the H$\alpha$ \cite{dickinson} template. We considered 70 frequency bands from 100 to 200\,MHz ($z=6$--13, spaced by $\Delta z=0.1$). 

The formalization of the component separation problem used in this and previous work \citep[e.g.,][]{chapman2012,chapman2013} models the 21\,cm signal from neutral hydrogen as noise, based on the low correlation between different redshift slices. We tested this hypothesis with our HI simulations and found that the correlation is still 10--20\% for a redshift separation of $\Delta z=0.1$ and can be neglected only for $\Delta z\geq0.2$. 

Another crucial aspect for EoR component separation is the complexity of the frequency behaviour of the foregrounds and our ability to model them accurately. In our case we focussed on the synchrotron emission, which is by far the dominant foreground component. We considered four synchrotron models to test the effect of different non-idealities in the frequency spectrum: spatial variations, curvature, and the presence of non-smooth features. Such effects are in principle very problematic for parametric methods such as the CCA. However, when modelling the signal as a power-law, we still get a good recovery of the spectral index. There is a small random error ($\Delta \beta_{\rm s}=0.01$--0.05 depending on the foreground simulation) and, for the model with curved spectrum a systematic error of 0.15. 

This accuracy in the synchrotron spectrum reduces the foreground contamination by several orders of magnitude when using a standard method to reconstruct the foreground maps and subtract them from the data. However, the cleaning is not always sufficient for the measurement of the tiny 21\,cm signal. The results can be improved substantially by modifying the foreground reconstruction and subtraction methods to make them robust against model errors. With these modifications, we finally obtain a very good cleaning of the cosmological signal for a wide range of frequencies (110--150\,MHz, $z=12$--8.5) and scales ($\log(k)\geq 1.5$\,Mpc$^{-1}$) for all the considered foreground simulations.

This work showed that the CCA method is very promising for improving our knowledge of the foreground emission and cleaning the EoR signal from foreground contamination, also in the presence of random and systematic departures of the true spectra from the parametric models adopted. The next steps towards applying this method to real SKA data is to test it against a frequency-dependent resolution and in the presence of point-source contaminants, both resolved and as a background.

\section{Aknowledgements}
We thank the anonymous referee for useful suggestions that improved the paper. AB and MLB acknowledge support from the European Research Council under the EC FP7 grant number 280127. MLB also acknowledges support from an STFC Advanced/Halliday fellowship. We thank K. Grainge for providing the SKA specifications, and F. Abdalla and E. Chapman for useful comments and suggestions. 
\bibliographystyle{mn2e}
\bibliography{biblio}

\begin{thebibliography}{}

\bibitem[\protect\citeauthoryear{{Becker} \& {Bolton}}{{Becker} \&
  {Bolton}}{2013}]{becker2013}
{Becker} G.~D.,  {Bolton} J.~S.,  2013, \mnras, 436, 1023

\bibitem[\protect\citeauthoryear{{Bernardi}, {de Bruyn}, {Harker}, {Brentjens},
  {Ciardi}, {Jeli{\'c}}, {Koopmans}, {Labropoulos}, {Offringa}, {Pandey},
  {Schaye}, {Thomas}, {Yatawatta} \& {Zaroubi}}{{Bernardi}
  et~al.}{2010}]{bernardi2010}
{Bernardi} G.,  {de Bruyn} A.~G.,  {Harker} G.,  {Brentjens} M.~A.,  {Ciardi}
  B.,  {Jeli{\'c}} V.,  {Koopmans} L.~V.~E.,  {Labropoulos} P.,  {Offringa} A.,
   {Pandey} V.~N.,  {Schaye} J.,  {Thomas} R.~M.,  {Yatawatta} S.,    {Zaroubi}
  S.,  2010, \aap, 522, A67

\bibitem[\protect\citeauthoryear{{Bharadwaj} \& {Ali}}{{Bharadwaj} \&
  {Ali}}{2005}]{Bharadwaj2005}
{Bharadwaj} S.,  {Ali} S.~S.,  2005, \mnras, 356, 1519

\bibitem[\protect\citeauthoryear{{Bonaldi}, {Bedini}, {Salerno}, {Baccigalupi}
  \& {de Zotti}}{{Bonaldi} et~al.}{2006}]{bonaldi2006}
{Bonaldi} A.,  {Bedini} L.,  {Salerno} E.,  {Baccigalupi} C.,    {de Zotti} G.,
   2006, \mnras, 373, 271

\bibitem[\protect\citeauthoryear{{Bonaldi} \& {Ricciardi}}{{Bonaldi} \&
  {Ricciardi}}{2012}]{special}
{Bonaldi} A.,  {Ricciardi} S.,  2012, Advances in Astronomy, 2012

\bibitem[\protect\citeauthoryear{{Bonaldi}, {Ricciardi}, {Leach}, {Stivoli},
  {Baccigalupi} \& {de Zotti}}{{Bonaldi} et~al.}{2007}]{bonaldi2007}
{Bonaldi} A.,  {Ricciardi} S.,  {Leach} S.,  {Stivoli} F.,  {Baccigalupi} C.,
   {de Zotti} G.,  2007, \mnras, 382, 1791

\bibitem[\protect\citeauthoryear{{Broadbent}, {Osborne} \&
  {Haslam}}{{Broadbent} et~al.}{1989}]{broad1989}
{Broadbent} A.,  {Osborne} J.~L.,    {Haslam} C.~G.~T.,  1989, \mnras, 237, 381

\bibitem[\protect\citeauthoryear{{Chapman}, {Abdalla}, {Bobin}, {Starck},
  {Harker}, {Jeli{\'c}}, {Labropoulos}, {Zaroubi}, {Brentjens}, {de Bruyn} \&
  {Koopmans}}{{Chapman} et~al.}{2013}]{chapman2013}
{Chapman} E.,  {Abdalla} F.~B.,  {Bobin} J.,  {Starck} J.-L.,  {Harker} G.,
  {Jeli{\'c}} V.,  {Labropoulos} P.,  {Zaroubi} S.,  {Brentjens} M.~A.,  {de
  Bruyn} A.~G.,    {Koopmans} L.~V.~E.,  2013, \mnras, 429, 165

\bibitem[\protect\citeauthoryear{{Chapman}, {Abdalla}, {Harker}, {Jeli{\'c}},
  {Labropoulos}, {Zaroubi}, {Brentjens}, {de Bruyn} \& {Koopmans}}{{Chapman}
  et~al.}{2012}]{chapman2012}
{Chapman} E.,  {Abdalla} F.~B.,  {Harker} G.,  {Jeli{\'c}} V.,  {Labropoulos}
  P.,  {Zaroubi} S.,  {Brentjens} M.~A.,  {de Bruyn} A.~G.,    {Koopmans}
  L.~V.~E.,  2012, \mnras, 423, 2518

\bibitem[\protect\citeauthoryear{{Chapman}, {Bonaldi}, {Harker}, {Jeli{\'c}},
  {Abdalla} et~al.,}{{Chapman} et~al.}{2014}]{chapman_SC}
{Chapman} E.,  {Bonaldi} A.,  {Harker} G.,  {Jeli{\'c}} V.,  {Abdalla} F.,
  et~al., 2014, in proceedings of "Advancing Astrophysics with the Square
  Kilometre Array" PoS(AASKA14), in press

\bibitem[\protect\citeauthoryear{{Di Matteo}, {Ciardi} \& {Miniati}}{{Di
  Matteo} et~al.}{2004}]{dimatteo2004}
{Di Matteo} T.,  {Ciardi} B.,    {Miniati} F.,  2004, \mnras, 355, 1053

\bibitem[\protect\citeauthoryear{{Di Matteo}, {Perna}, {Abel} \& {Rees}}{{Di
  Matteo} et~al.}{2002}]{dimatteo2002}
{Di Matteo} T.,  {Perna} R.,  {Abel} T.,    {Rees} M.~J.,  2002, \apj, 564, 576

\bibitem[\protect\citeauthoryear{{Dickinson}, {Davies} \& {Davis}}{{Dickinson}
  et~al.}{2003}]{dickinson}
{Dickinson} C.,  {Davies} R.~D.,    {Davis} R.~J.,  2003, \mnras, 341, 369

\bibitem[\protect\citeauthoryear{{Dillon}, {Liu}, {Williams}, {Hewitt},
  {Tegmark} et~al.,}{{Dillon} et~al.}{2013}]{dillon2013}
{Dillon} J.~S.,  {Liu} A.,  {Williams} C.~L.,  {Hewitt} J.~N.,  {Tegmark} M.,
   et~al., 2013, ArXiv e-prints

\bibitem[\protect\citeauthoryear{{Draine}}{{Draine}}{2011}]{draine2011}
{Draine} B.~T.,  2011, {Physics of the Interstellar and Intergalactic Medium}

\bibitem[\protect\citeauthoryear{{Fan}, {Strauss}, {Becker}, {White}, {Gunn},
  {Knapp}, {Richards}, {Schneider}, {Brinkmann} \& {Fukugita}}{{Fan}
  et~al.}{2006}]{fan2006}
{Fan} X.,  {Strauss} M.~A.,  {Becker} R.~H.,  {White} R.~L.,  {Gunn} J.~E.,
  {Knapp} G.~R.,  {Richards} G.~T.,  {Schneider} D.~P.,  {Brinkmann} J.,
  {Fukugita} M.,  2006, \aj, 132, 117

\bibitem[\protect\citeauthoryear{{Furlanetto}, {Oh} \& {Briggs}}{{Furlanetto}
  et~al.}{2006}]{furlanetto2006}
{Furlanetto} S.~R.,  {Oh} S.~P.,    {Briggs} F.~H.,  2006, \physrep, 433, 181

\bibitem[\protect\citeauthoryear{{Furlanetto}, {Zaldarriaga} \&
  {Hernquist}}{{Furlanetto} et~al.}{2004}]{FZH2004}
{Furlanetto} S.~R.,  {Zaldarriaga} M.,    {Hernquist} L.,  2004, \apj, 613, 1

\bibitem[\protect\citeauthoryear{{Gleser}, {Nusser} \& {Benson}}{{Gleser}
  et~al.}{2008}]{gleser2008}
{Gleser} L.,  {Nusser} A.,    {Benson} A.~J.,  2008, \mnras, 391, 383

\bibitem[\protect\citeauthoryear{{G{\'o}rski}, {Hivon}, {Banday}, {Wandelt},
  {Hansen}, {Reinecke} \& {Bartelmann}}{{G{\'o}rski} et~al.}{2005}]{gorski}
{G{\'o}rski} K.~M.,  {Hivon} E.,  {Banday} A.~J.,  {Wandelt} B.~D.,  {Hansen}
  F.~K.,  {Reinecke} M.,    {Bartelmann} M.,  2005, \apj, 622, 759

\bibitem[\protect\citeauthoryear{{Harker}, {Zaroubi}, {Bernardi}, {Brentjens},
  {de Bruyn}, {Ciardi}, {Jeli{\'c}}, {Koopmans}, {Labropoulos}, {Mellema},
  {Offringa}, {Pandey}, {Schaye}, {Thomas} \& {Yatawatta}}{{Harker}
  et~al.}{2009}]{harker2009}
{Harker} G.,  {Zaroubi} S.,  {Bernardi} G.,  {Brentjens} M.~A.,  {de Bruyn}
  A.~G.,  {Ciardi} B.,  {Jeli{\'c}} V.,  {Koopmans} L.~V.~E.,  {Labropoulos}
  P.,  {Mellema} G.,  {Offringa} A.,  {Pandey} V.~N.,  {Schaye} J.,  {Thomas}
  R.~M.,    {Yatawatta} S.,  2009, \mnras, 397, 1138

\bibitem[\protect\citeauthoryear{{Haslam}, {Salter}, {Stoffel} \&
  {Wilson}}{{Haslam} et~al.}{1982}]{haslam}
{Haslam} C.~G.~T.,  {Salter} C.~J.,  {Stoffel} H.,    {Wilson} W.~E.,  1982,
  \aaps, 47, 1

\bibitem[\protect\citeauthoryear{{Hinshaw}, {Larson}, {Komatsu}, {Spergel},
  {Bennett}, {Dunkley} et~al.,}{{Hinshaw} et~al.}{2013}]{wmap9-cosmo}
{Hinshaw} G.,  {Larson} D.,  {Komatsu} E.,  {Spergel} D.~N.,  {Bennett} C.~L.,
  {Dunkley} J.,    et~al., 2013, \apjs, 208, 19

\bibitem[\protect\citeauthoryear{{Jeli{\'c}}, {Zaroubi}, {Labropoulos},
  {Bernardi}, {de Bruyn} \& {Koopmans}}{{Jeli{\'c}} et~al.}{2010}]{jelic2010}
{Jeli{\'c}} V.,  {Zaroubi} S.,  {Labropoulos} P.,  {Bernardi} G.,  {de Bruyn}
  A.~G.,    {Koopmans} L.~V.~E.,  2010, \mnras, 409, 1647

\bibitem[\protect\citeauthoryear{{Jeli{\'c}}, {Zaroubi}, {Labropoulos},
  {Thomas}, {Bernardi}, {Brentjens}, {de Bruyn}, {Ciardi}, {Harker},
  {Koopmans}, {Pandey}, {Schaye} \& {Yatawatta}}{{Jeli{\'c}}
  et~al.}{2008}]{jelic2008}
{Jeli{\'c}} V.,  {Zaroubi} S.,  {Labropoulos} P.,  {Thomas} R.~M.,  {Bernardi}
  G.,  {Brentjens} M.~A.,  {de Bruyn} A.~G.,  {Ciardi} B.,  {Harker} G.,
  {Koopmans} L.~V.~E.,  {Pandey} V.~N.,  {Schaye} J.,    {Yatawatta} S.,  2008,
  \mnras, 389, 1319

\bibitem[\protect\citeauthoryear{{Liu} \& {Tegmark}}{{Liu} \&
  {Tegmark}}{2011}]{liu2011}
{Liu} A.,  {Tegmark} M.,  2011, \prd, 83, 103006

\bibitem[\protect\citeauthoryear{{Liu} \& {Tegmark}}{{Liu} \&
  {Tegmark}}{2012}]{liu2012}
{Liu} A.,  {Tegmark} M.,  2012, \mnras, 419, 3491

\bibitem[\protect\citeauthoryear{{Liu}, {Tegmark}, {Bowman}, {Hewitt} \&
  {Zaldarriaga}}{{Liu} et~al.}{2009}]{liu2009}
{Liu} A.,  {Tegmark} M.,  {Bowman} J.,  {Hewitt} J.,    {Zaldarriaga} M.,
  2009, \mnras, 398, 401

\bibitem[\protect\citeauthoryear{{Mellema}, {Iliev}, {Pen} \&
  {Shapiro}}{{Mellema} et~al.}{2006}]{mellema2006}
{Mellema} G.,  {Iliev} I.~T.,  {Pen} U.-L.,    {Shapiro} P.~R.,  2006, \mnras,
  372, 679

\bibitem[\protect\citeauthoryear{{Mellema}, {Koopmans}, {Abdalla}, {Bernardi},
  {Ciardi}, {Daiboo}, {de Bruyn} et~al.,}{{Mellema} et~al.}{2013}]{mellema2013}
{Mellema} G.,  {Koopmans} L.~V.~E.,  {Abdalla} F.~A.,  {Bernardi} G.,  {Ciardi}
  B.,  {Daiboo} S.,  {de Bruyn} A.~G.,    et~al., 2013, Experimental Astronomy,
  36, 235

\bibitem[\protect\citeauthoryear{{Mesinger}, {Furlanetto} \& {Cen}}{{Mesinger}
  et~al.}{2011}]{mesinger2011}
{Mesinger} A.,  {Furlanetto} S.,    {Cen} R.,  2011, \mnras, 411, 955

\bibitem[\protect\citeauthoryear{{Moore}, {Aguirre}, {Parsons}, {Jacobs} \&
  {Pober}}{{Moore} et~al.}{2013}]{moore2013}
{Moore} D.~F.,  {Aguirre} J.~E.,  {Parsons} A.~R.,  {Jacobs} D.~C.,    {Pober}
  J.~C.,  2013, \apj, 769, 154

\bibitem[\protect\citeauthoryear{{Morales}, {Bowman} \& {Hewitt}}{{Morales}
  et~al.}{2006}]{morales2006}
{Morales} M.~F.,  {Bowman} J.~D.,    {Hewitt} J.~N.,  2006, \apj, 648, 767

\bibitem[\protect\citeauthoryear{{Morales} \& {Wyithe}}{{Morales} \&
  {Wyithe}}{2010}]{morales2010}
{Morales} M.~F.,  {Wyithe} J.~S.~B.,  2010, \araa, 48, 127

\bibitem[\protect\citeauthoryear{{Oh} \& {Mack}}{{Oh} \& {Mack}}{2003}]{oh2003}
{Oh} S.~P.,  {Mack} K.~J.,  2003, \mnras, 346, 871

\bibitem[\protect\citeauthoryear{{Paciga}, {Albert}, {Bandura}, {Chang},
  {Gupta}, {Hirata}, {Odegova}, {Pen}, {Peterson}, {Roy}, {Shaw}, {Sigurdson}
  \& {Voytek}}{{Paciga} et~al.}{2013}]{paciga2013}
{Paciga} G.,  {Albert} J.~G.,  {Bandura} K.,  {Chang} T.-C.,  {Gupta} Y.,
  {Hirata} C.,  {Odegova} J.,  {Pen} U.-L.,  {Peterson} J.~B.,  {Roy} J.,
  {Shaw} J.~R.,  {Sigurdson} K.,    {Voytek} T.,  2013, \mnras, 433, 639

\bibitem[\protect\citeauthoryear{{Parsons}, {Liu}, {Aguirre}, {Ali}, {Bradley}
  et~al.,}{{Parsons} et~al.}{2013}]{parsons2013}
{Parsons} A.~R.,  {Liu} A.,  {Aguirre} J.~E.,  {Ali} Z.~S.,  {Bradley} R.~F.,
   et~al., 2013, ArXiv e-prints

\bibitem[\protect\citeauthoryear{{Petrovic} \& {Oh}}{{Petrovic} \&
  {Oh}}{2011}]{petrovic2011}
{Petrovic} N.,  {Oh} S.~P.,  2011, \mnras, 413, 2103

\bibitem[\protect\citeauthoryear{{Planck Collaboration}}{{Planck
  Collaboration}}{2013}]{gouldbelt}
{Planck Collaboration} 2013, \aap, 557, A53

\bibitem[\protect\citeauthoryear{{Planck Collaboration}, {Ade}, {Aghanim},
  {Armitage-Caplan}, {Arnaud}, {Ashdown}, {Atrio-Barandela}, {Aumont},
  {Baccigalupi}, {Banday} \& et al.}{{Planck Collaboration}
  et~al.}{2013}]{planck-cosmo}
{Planck Collaboration} {Ade} P.~A.~R.,  {Aghanim} N.,  {Armitage-Caplan} C.,
  {Arnaud} M.,  {Ashdown} M.,  {Atrio-Barandela} F.,  {Aumont} J.,
  {Baccigalupi} C.,  {Banday} A.~J.,    et al. 2013, ArXiv e-prints 1303.5076

\bibitem[\protect\citeauthoryear{{Pritchard} \& {Loeb}}{{Pritchard} \&
  {Loeb}}{2012}]{pritchard2012}
{Pritchard} J.~R.,  {Loeb} A.,  2012, Reports on Progress in Physics, 75,
  086901

\bibitem[\protect\citeauthoryear{{Remazeilles}, {Dickinson}, {Banday}, M.-A. \&
  {Ghosh}}{{Remazeilles} et~al.}{2014}]{remazeilles2014}
{Remazeilles} M.,  {Dickinson} C.,  {Banday} A.~J.,  M.-A. B.,    {Ghosh} T.,
  2014, ArXiv e-prints 1411.3628

\bibitem[\protect\citeauthoryear{{Ricciardi}, {Bonaldi}, {Natoli}, {Polenta},
  {Baccigalupi}, {Salerno}, {Kayabol}, {Bedini} \& {de Zotti}}{{Ricciardi}
  et~al.}{2010}]{ricciardi2010}
{Ricciardi} S.,  {Bonaldi} A.,  {Natoli} P.,  {Polenta} G.,  {Baccigalupi} C.,
  {Salerno} E.,  {Kayabol} K.,  {Bedini} L.,    {de Zotti} G.,  2010, \mnras,
  406, 1644

\bibitem[\protect\citeauthoryear{{Santos}, {Cooray} \& {Knox}}{{Santos}
  et~al.}{2005}]{santos2005}
{Santos} M.~G.,  {Cooray} A.,    {Knox} L.,  2005, \apj, 625, 575

\bibitem[\protect\citeauthoryear{{Stolyarov}, {Hobson}, {Lasenby} \&
  {Barreiro}}{{Stolyarov} et~al.}{2005}]{stolyarov2005}
{Stolyarov} V.,  {Hobson} M.~P.,  {Lasenby} A.~N.,    {Barreiro} R.~B.,  2005,
  \mnras, 357, 145

\bibitem[\protect\citeauthoryear{{Strong}, {Orlando} \& {Jaffe}}{{Strong}
  et~al.}{2011}]{galprop}
{Strong} A.~W.,  {Orlando} E.,    {Jaffe} T.~R.,  2011, \aap, 534, A54

\bibitem[\protect\citeauthoryear{{Yatawatta}, {de Bruyn}, {Brentjens},
  {Labropoulos}, {Pandey}, {Kazemi}, {Zaroubi}, {Koopmans} et~al.,}{{Yatawatta}
  et~al.}{2013}]{yatawatta2013}
{Yatawatta} S.,  {de Bruyn} A.~G.,  {Brentjens} M.~A.,  {Labropoulos} P.,
  {Pandey} V.~N.,  {Kazemi} S.,  {Zaroubi} S.,  {Koopmans} L.~V.~E.,    et~al.,
  2013, \aap, 550, A136

\bibitem[\protect\citeauthoryear{{Zahn}, {Reichardt}, {Shaw}, {Lidz}, {Aird},
  {Benson}, {Bleem}, {Carlstrom}, {Chang}, {Cho} et~al.,}{{Zahn}
  et~al.}{2012}]{zahn2012}
{Zahn} O.,  {Reichardt} C.~L.,  {Shaw} L.,  {Lidz} A.,  {Aird} K.~A.,  {Benson}
  B.~A.,  {Bleem} L.~E.,  {Carlstrom} J.~E.,  {Chang} C.~L.,  {Cho} H.~M.,
  et~al., 2012, \apj, 756, 65

\end{thebibliography}
\end{document}